\documentclass[aps,prm,amsmath,amssymb,reprint,superscriptaddress,nofootinbib]{revtex4-2}
\usepackage{charter,graphicx,verbatim,threeparttable,float,amssymb,gensymb }
\usepackage{upgreek}
\usepackage{booktabs}
\usepackage{tabularx}

\newcommand{\super}{\textit{Ln}$_2$Ti$_9$Sb$_{11}$}
\newcommand{\LAsuper}{La$_2$Ti$_9$Sb$_{11}$}
\newcommand{\NDsuper}{Nd$_2$Ti$_9$Sb$_{11}$}
\newcommand{\CEsuper}{Ce$_2$Ti$_9$Sb$_{11}$}
\newcommand{\PRsuper}{Pr$_2$Ti$_9$Sb$_{11}$}
\newcommand{\disord}{\textit{Ln}$_2$Ti$_7$Sb$_{12}$}
\begin{document}

\preprint{APS/123-QED}

\title{Isolated spin ladders in  \textit{Ln}$_2$Ti$_9$Sb$_{11}$ (\textit{Ln:}La--Nd) metals}

\author{Brenden R. Ortiz}
\email{ortizbr@ornl.gov}
\affiliation{Materials Science and Technology Division, Oak Ridge National Laboratory, Oak Ridge, TN 37831, USA}

\author{Heda Zhang} 
\affiliation{Materials Science and Technology Division, Oak Ridge National Laboratory, Oak Ridge, TN 37831, USA}

\author{Karolina G\'{o}rnicka} 
\affiliation{Materials Science and Technology Division, Oak Ridge National Laboratory, Oak Ridge, TN 37831, USA}
\affiliation{Applied Physics and Mathematics Department, Advanced Materials Centre, Gdansk University of Technology, ul. Narutowicza 11/12, 80-233 Gdansk, Poland}

\author{Matthew S. Cook} 
\affiliation{Materials Science and Technology Division, Oak Ridge National Laboratory, Oak Ridge, TN 37831, USA}

\author{Suchismita Sarker} 
\affiliation{Cornell High Energy Synchrotron Source, Cornell University, Ithaca, New York 14853, USA}

\author{Satoshi Okamoto} 
\affiliation{Materials Science and Technology Division, Oak Ridge National Laboratory, Oak Ridge, TN 37831, USA}

\author{Jiaqiang Yan} 
\affiliation{Materials Science and Technology Division, Oak Ridge National Laboratory, Oak Ridge, TN 37831, USA}

\date{\today}
\begin{abstract}
Here we present the discovery and characterization of a series of antimonides \super\ (\textit{Ln}: La--Nd) which exhibit well-isolated, $n=2$ rare-earth spin ladders. We discuss the structure of the new compounds, with a particular focus on the magnetic \textit{Ln} spin ladders. Nd$_2$Ti$_9$Sb$_{11}$ and Ce$_2$Ti$_9$Sb$_{11}$ exhibit antiferromagnetic interactions and a well-defined doublet ground state, whereas Pr$_2$Ti$_9$Sb$_{11}$ exhibits a weakly magnetic singlet ground state. Nd$_2$Ti$_9$Sb$_{11}$ is a poor metal with an electrical resistivity of 0.1m$\Omega$-cm at 300~K and weak temperature dependence. The thermal conductivity along the ladder exhibits significant field dependence even at 40\,K, considerably higher than the magnetic ordering temperature of 1.1\,K. Compared to compounds with transition metal spin ladders, the rare-earth elements impart much lower energy scales, making these compounds highly tunable with external stimuli like magnetic fields. The diverse magnetism of the rare-earth ions and RKKY interactions further contribute to the potential for a wide array of rich magnetic ground states, positioning these materials as a rare example of an inorganic square spin-ladder platform.
\end{abstract}
\maketitle
\section{Introduction}

Spin-chain systems serve as one of the fundamental platforms for exploring condensed matter physics, offering profound insights into quantum magnetism and low-dimensional phenomena. An extraordinary number of theoretical predictions have been realized in spin-chain platforms. From Haldane's prediction that a one-dimensional antiferromagnet with Heisenberg anisotropy has an energy gap for integral $S$ and no gap for half-integral $S$,\cite{haldane1983nonlinear} to the analytical solutions offered by the Bethe ansatz,\cite{bethe1931theorie,hulthen1938austauschproblem} much insight has been gleaned from a relatively simple model system. These and subsequent theoretical efforts\cite{sutherland1970two,baxter1971one,johnson1973vertical,klumper1993thermodynamics} have led to a wide array of developments in solid-state physics, including development of spin-wave theory\cite{anderson1952approximate,anderson1951limits,kubo1952spin} and a wide variety of experimental and theoretical efforts towards understanding the nuances of spin-chain systems.

Building on the one-dimensional spin chain, a natural extension is to consider a system of $n$ coupled chains. Dubbed $n$-leg ``spin ladders,'' these systems bridge the gap between the 1D spin chain and a 2D square lattice. As such, spin ladders provide a unique platform to study the evolution of quantum magnetism entangled with the dimensional crossover from 1D to 2D. A full theoretical treatment of the $n$-leg spin ladder is not available, though a myriad of theoretical predictions have proven quite robust. For example, the predicted ground state of a $S=1/2$ ladder is a gapped singlet spin liquid when $n$ is even, and a gapless singlet spin liquid when $n$ is odd.\cite{dagotto1996surprises} The emergence of superconductivity in doped, even-$n$ ladders after suppression of antiferromagnetism is also a poignant prediction from spin-ladder theory.\cite{dagotto1999experiments,kageyama2008spin}

Perhaps the most ubiquitous example of a \( n=2 \) spin-ladder  is the generalized set of strontium cuprates Sr$_{n-1}$Cu$_{n}$O$_{2n-1}$, which can even be tuned to exhibit different dimensionalities of \( n \)-legged spin ladders.\cite{hiroi1991new} Several other examples of inorganic spin-ladder structure types are CaV$_2$O$_5$,\cite{koo1999analysis} Sr$_3$Fe$_2$O$_5$,\cite{kageyama2008spin} VO$_2$(P$_2$O$_7$),\cite{gorbunova1979structure} and BaFe$_2$S$_3$.\cite{takahashi2015pressure} There are also a surprising number of complex organic compounds that exhibit spin ladders as well, including Cu$_2$C$_{10}$H$_{24}$N$_4$Cl$_4$,\cite{chiari1990exchange} (C$_{5}$H$_{12}$N)Cu$_{2}$Br$_{3}$,\cite{komm2012c5h12n} (C$_{4}$H$_{14}$N$_{2}$)Cu$_{2}$\textit{X}$_{6}$ (\textit{X}=Cl,Br),\cite{biswal2023crystal} [Ph(NH$_{3}$)-[18]crown-6][Ni(dmit)$_{2}$],\cite{akutagawa2002supramolecular} [(DT-TTF)$_{2}$][Au(mnt)$_{2}$],\cite{rovira1997organic} and C$_{28}$H$_{42}$N$_{4}$O$_{4}$ (BIP-TENO).\cite{nomura2022metastable} 

Many of the most well-known spin ladders build on $S=1/2$ Cu, owing to the wealth of theoretical and computational predictions available for $S=1/2$ ladders. It is not always known, \textit{ab initio}, what effects varying the spin will have on the magnetic ground state. However, having the ability to mix a larger variety of S (or J) with diverse anisotropies could result in a rich array of magnetic interactions.\cite{kageyama2008spin} Compounds utilizing rare-earth elements are one potential route towards highly tunable spin ladders, as a wide variety of effective $J$ and anisotropies can be accessed without changing the requisite chemistry. 

Efforts to develop rare-earth spin-ladder compounds have produced materials like \textit{Ln}B$_{44}$Si$_2$,\cite{higashi1997crystal,mori2006crystal} Ba$_2$\textit{Ln}$_2$Ge$_4$O$_{13}$,\cite{zhou2024structure} and Sr\textit{Ln}$_2$O$_4$.\cite{orlandi2025magnetic} However, all of these compounds require a generous interpretation of bond distances to qualify them as \( n=2 \) spin ladders. For example, in SrTb$_2$O$_4$ a \textit{Ln--Ln} distance of 3.6\AA~ yields a \( n=2 \) zig-zag ladder, but 3.9\AA~ yields 2D corrugated sheets, and 4\AA~ yields 3D tubes. Similarly, Ba$_2$\textit{Ln}$_2$Ge$_4$O$_{13}$ is probably more accurately described as a system of \textit{Ln--Ln} dimers (3.6\AA~), as the inter-dimer distance (5.4\AA) that creates the ladder is nearly the same as the inter-ladder distance (5.9\AA). \textit{Ln}B$_{44}$Si$_2$ suffers from a similar issue, as the nearest-neighbor interactions create rectangular plaquettes with sides of 3.9--4.4\AA, but the next-nearest neighbor distance creating the spin-ladder is nearly 5.2\AA. Thus, there is a distinct need for systems exhibiting well-isolated, well-defined spin ladders constructed from rare-earth elements.
 
In this work we present the single-crystal synthesis and characterization of a new spin-ladder family \super\ (\textit{Ln}: La, Ce, Pr, Nd). This family exhibits a \( n=2 \) spin ladder of \textit{Ln} interwoven into a complex network of Ti--Ti and Ti--Sb. The ladders approximate a square ladder with nearly equal leg and rung spacing (4.4--4.6\AA). The ladders are also well-isolated, as the ladder ''center-to-center'' distance is $>$10\AA. Here we examine the magnetic properties of the \super\ family, focusing on Nd$_2$Ti$_9$Sb$_{11}$ for additional electronic and thermal transport measurements. We observe that both Nd$_2$Ti$_9$Sb$_{11}$ and Ce$_2$Ti$_9$Sb$_{11}$ exhibit magnetic ground states with hallmark features of low-dimensional magnetism. Conversely, Pr$_2$Ti$_9$Sb$_{11}$ adopts a weakly magnetic singlet ground state owing to the complex coordination and non-Kramers nature of Pr$^{3+}$. Among inorganic spin-ladder compounds these materials stand apart as examples of well-isolated, square spin-ladder compounds built from rare-earth magnetism.

\section{Methods}
\subsection{Synthesis}
\super\ single crystals are grown from metallic Sn using the flux method. Rare-earth metal pieces (Ames) were placed with Ti powder (Alfa 99.9\%), Sb shot (Alfa 99.999\%), and Sn shot (Alfa 99.9\%) at a ratio of 0.25:0.75:4:6 \textit{Ln}:Ti:Sb:Sn into 2~mL Canfield crucibles fitted with a catch crucible and a porous frit.\cite{canfield2016use} The crucibles were sealed under approximately 0.7~atm of argon gas in fused silica ampoules. Samples were heated to 1100\degree C at a rate of 200\degree C/hr and thermalized at 1100\degree C for 12~h before cooling to 950\degree C at a rate of 2\degree C/hr. Excess flux was removed from the crystals by centrifugation at 950\degree C. Resulting crystals are brittle polyhedral rods with a silver luster and complex faceting. Crystals are stable in air, water, and common solvents.

Polycrystalline samples can be produced through mechanochemical methods. \textit{Ln} shavings filed from rods (Ames) were placed with Ti powder (Alfa 99.9\%) and Sb shot (Alfa 99.999\%) into a tungsten carbide SPEX ball-mill vial and milled in a SPEX 8000M mixer/mill for 90~m. The resulting precursor powder was extracted, sealed under argon, and annealed at 950\degree C for 48~h. Please note that polycrystalline measurements in the manuscript are produced from ground single crystals. We mention the synthesis of the polycrystalline samples from direct reaction to emphasize that the pure antimonide is a stable phase in the absence of Sn.

\subsection{Bulk Characterization}
Single crystals of \super\ were mounted on kapton loops with Paratone oil for single crystal x-ray diffraction (SCXRD). Diffraction data were collected at $\approx$100~K on a Bruker D8 Advance Quest diffractometer with a graphite monochromator using Mo K$\alpha$ radiation ($\lambda$ = 0.71073~\AA). Data integration, reduction, and structure solution was performed using the Bruker APEX4 software package. All atoms were refined with anisotropic thermal parameters. CIF files and associated crystallographic tables for all structures are included in the ESI.\cite{ESI} High-resolution precession images were performed at the Cornell High Energy X-ray Synchrotron Source (CHESS), beamline IDB4-QM2 (45keV). Elemental analysis was carried out on as-grown crystals using a Hitachi-TM3000 microscope equipped with a Bruker Quantax 70 EDS system. 

Magnetization measurements (300 -- 1.8~K) on crystals of \super\ were performed in a 7~T Quantum Design Magnetic Property Measurement System (MPMS3) SQUID magnetometer in vibrating-sample magnetometry (VSM) mode. Additional measurements (1.8 -- 0.4~K) utilized the Quantum Design iHe-3 He$^{3}$ insert for the MPMS3. Data from the iHe-3 He$^{3}$ system is matched to the MPMS3 data over the range 4--1.8~K to create continuous magnetization data sets from 300--0.4~K.

Electronic transport measurements were performed in a Quantum Design 12~T Dynacool PPMS. A naturally faceted single crystal was selected and oriented by performing facet scans on a PANalytical Xpert Pro diffactometer ($\lambda$ = 1.5406~\AA). The sample was polished into a resistivity bar using a Struers AccuStop sample holder and Crystalbond 509. The bar was mounted to a sheet of sapphire, which was subsequently adhered to the sample puck stage using GE varnish. The sample surface was cleaned by etching with an Ar plasma before applying a thin layer of Au. Contact to the Au pads was then made using silver paint (DuPont cp4929N-100) and platinum wire (Alfa, 0.05~mm Premion 99.995\%) in a four-wire geometry for magnetoresistance measurements. Transport measurements utilized a 2.5~mA probe current.

The same crystal from electronic transport measurements was mounted for thermal transport measurements on a custom-built puck compatible with the Quantum Design PPMS cryostat. A Keithley 6221 current source is used to supply a current through a 1 k$\Omega$ chip resistor and calculate the thermal power (P) applied to the sample through Joule heating. The temperature difference ($\Delta$T) induced by this thermal current is obtained by a two-point measurement through Cernox sensors attached to the sample \textit{via} gold wires and monitored by a Lakeshore 336 temperature controller. The thermal conductivity is calculated based on $\kappa_{xx}$=(\textit{l}/A)(P/$\Delta$T), where \textit{l} is the distance between two sensing points on the sample, $\Delta$T is the corresponding temperature difference, A is the cross-sectional area of the sample, and P is the power supplied by the heater. 

Heat capacity measurements (300--1.8~K) were performed in a Quantum Design 9~T Dynacool Physical Property Measurement System (PPMS). Additional heat capacity measurements (3.5--0.1~K) utilized the Quantum Design Dilution Refrigerator insert for the Quantum Design 9~T Dynacool PPMS. Thermal contact was achieved using Apezion N-grease.  

\subsection{Electronic Structure}

In an effort to describe the underlying electronic structure of the \super compounds, we performed first-principles density functional theory calculations on \LAsuper~ in a non-spin-polarized configuration with a full 4f shell and no magnetic character. We use the Vienna {\it ab}-initio simulation package (VASP) \cite{kresse1996efficient,kresse1996efficiency} based on the projector-augmented wave method \cite{kresse1999ultrasoft} with the generalized gradient approximation in the parametrization of Perdew, Burke, and Enzerhof \cite{perdew1996}
the for exchange–correlation functional.  
For La and Sb, the standard potentials were used. For
Ti, the $s$ states were treated as valence states (Ti$_{\rm sv}$ in the VASP distribution). We use a $4 \times 4 \times 4$ k-point grid and an energy cutoff of 500~eV. We did not consider a $+U$ correction or spin-orbit coupling. To generate the primitive unit cell based on our structural data, we use SeeK-path.\cite{hinuma2017band,togo2024spglib}

\section{Results and Discussion}
\subsection{Synthesis and Crystal Structure}

During attempts to prepare analogs of the recently reported \textit{Ln}V$_3$Sb$_4$\cite{ortiz2023ybv} and \textit{Ln}Ti$_3$Bi$_4$\cite{ortiz2019new,ortiz2023evolution,ovchinnikov2018synthesis,ovchinnikov2019bismuth,motoyama2018magnetic,chen2023134,guo2023134,ortiz2024intricate,guo20241,cheng2024giant} compounds, we identified the presence of large polyhedral crystals that form as a competing phase for large rare-earth species (\textit{Ln}: La--Nd). The new compounds have a narrow thermal stability in Sn flux, and cooling below 950\degree C causes the crystals to decompose into binary phases. The pure antimonide compound can be prepared from solid-state methods (see methods section), but we were unable to prepare single crystals through an antimony flux. Crystals prepared from a Sn-flux contain a negligible amount of Sn ($\leq2$at.$\%$) and can grow large (2 $\times$ 2 $\times$ 10~mm) pseudo-octagonal rods with good faceting. 

\begin{figure}
\includegraphics[width=\linewidth]{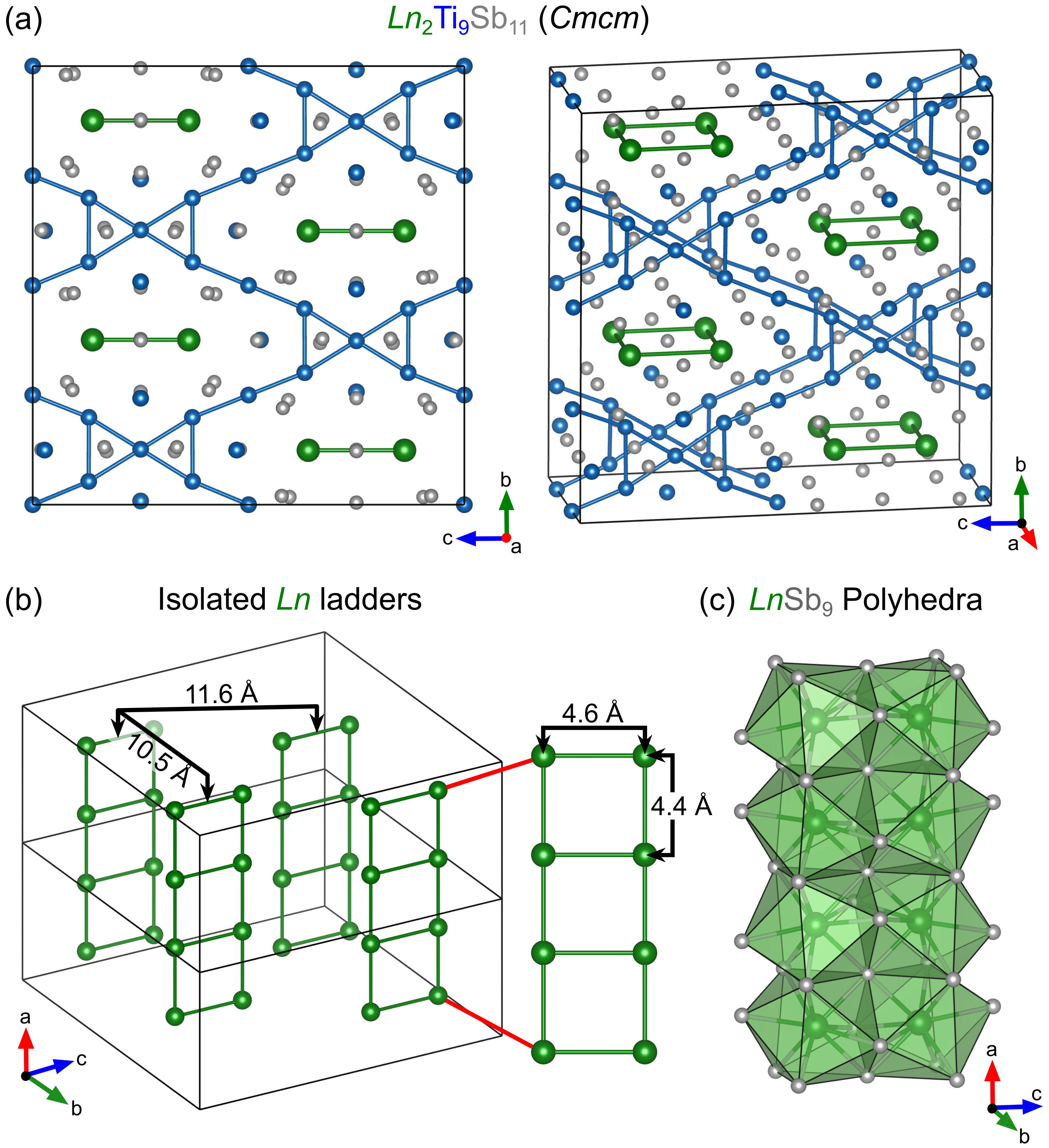}
\caption{(a) Two projections of the \super\ structure, highlighting the Ti--Ti and \textit{Ln}--\textit{Ln} interactions. The structure is complex, but can be understood as a \( n=2 \) spin ladder of \textit{Ln} that propagates parallel to \textit{a}. The ladders are interwoven with a complex Ti--Ti and Ti--Sb network in the \textit{bc} plane. (b) The ladders in \super\ are well-isolated from each other and are a close approximation to an ideal square ladder. (c) \textit{Ln} atoms are 9-fold coordinated by Sb (d$\sim3.5$\AA) isolating the \textit{Ln} sublattice from the Ti--Ti network.}
\label{fig:xtal}
\end{figure}

\begin{figure*}
\includegraphics[width=1\textwidth]{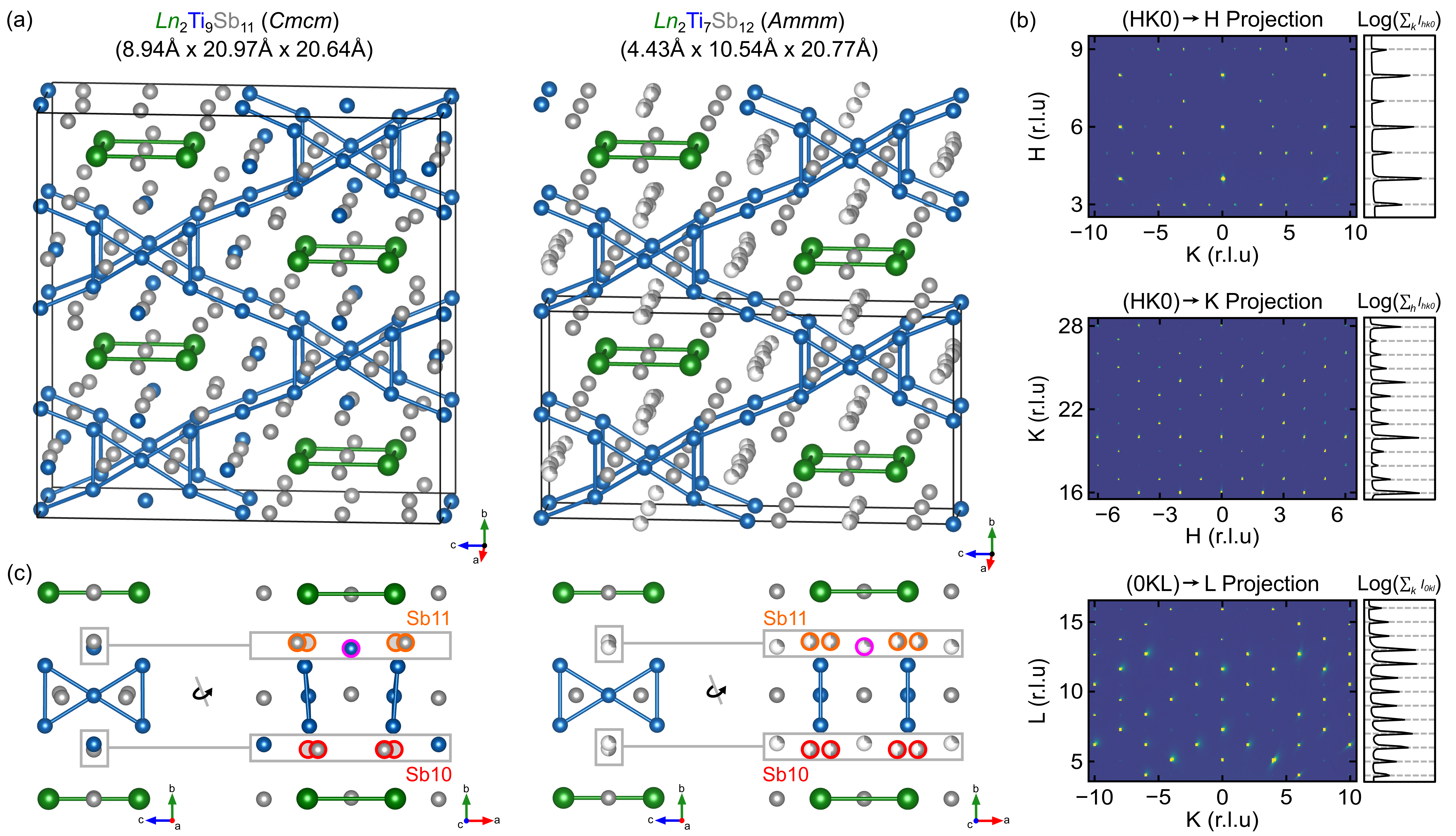}
\caption{(a) Unit cell comparison between our newly discovered \super\ and the previously reported \disord. For ease of comparison, we have converted the literature result into the equivalent \textit{Ammm} unit cell, aligning the \textit{Ln} and Ti networks. It is readily apparent that \super\ is approximately an ordered 2$\times$2$\times$1 supercell of disordered \disord. (b) Precession images of Nd$_2$Ti$_9$Sb$_{11}$ indexed in the larger 2$\times$2$\times$1 \textit{Cmcm} supercell. Each image includes a projection of the reflections onto one axis (e.g. HK0 projected onto H) to highlight that the larger cell captures every integer reflection. (c) Comparison between the two structures, highlighting the differences on the Sb10 (red highlight) and Sb11 (orange highlight) split-sites. Also note the fully-occupied Ti (purple highlight) in \super\ compared to the split Sb site in \disord.}
\label{fig:super}
\end{figure*}

The new compound indexes to a orthorhombic (\textit{Cmcm}) unit cell with lattice parameters (\textit{a},\textit{b},\textit{c}): (8.9\AA, 20.9\AA, 20.6\AA). Single-crystal diffraction (SCXRD) results propose the formula \super, which is in excellent agreement with energy-dispersive spectroscopy (EDS) on clean fracture surfaces (9(1)\% Ln, 42(2)\% Ti, and 49(2)\% Sb). Figure \ref{fig:xtal}(a,b) shows the refined crystal structure of  \super\ with titanium (blue) Ti--Ti bonds drawn for $d\leq3$\AA~ and rare-earth (green) \textit{Ln}--\textit{Ln} interactions highlighted with $d\leq5$\AA. For visual clarity we have reduced the display size of the Sb atoms (gray) and have also omitted any \textit{X}-Sb bonds. CIF files for all \super\ compounds presented here have been included in the ESI.\cite{ESI}

The complex crystal structure of \super\ exhibits two key motifs. The Ti sublattice is a highly distorted chain-like network which contains remnants of the kagome lattice that forms in the neighboring \textit{AM}$_3$\textit{X}$_4$ phases. The hallmark feature of the \super\ structure is the \textit{Ln} sublattice, shown in Figure \ref{fig:xtal}(b). Despite the complex structure, there is only a single \textit{Ln} site in the paramagnetic (high-temperature) structure, which generates well-isolated \( n=2 \) spin-ladders. The rung and leg spacings are very similar (4.4--4.6\AA), representing one of the closest approximations of a square spin ladder. The ladders are also well-isolated from one another, with a ''center-to-center'' distance of approximately 10.5\AA. \textit{Ln} atoms are coordinated by 9 Sb atoms with an average \textit{Ln}-Sb distance of 3.5\AA, structurally isolating them from the Ti-Ti network (Figure \ref{fig:xtal}(c)).

The structural motifs in \super\ are highly reminiscent of the previously reported \disord\ (Bie et al.).\cite{bie2007ternary} Figure \ref{fig:super} provides a graphical comparison between our results and the literature results from Bie et al. For ease of visual comparison we have shifted both the origin and the axes (\textit{Cmmm} to \textit{Ammm}) for the literature result. The Ti and \textit{Ln} networks are obviously similar between the two families, though the model proposed by Bie et al. includes numerous split Sb sites.\cite{bie2007ternary} Our compound is nearly ordered, leading to substantial differences in the indexed cell from diffraction. There is also a substantial difference in the reported compositions. Our spectroscopic results suggest that the samples are 9\% \textit{Ln}, 42\% Ti, and 49\% Sb, which is in excellent agreement with single crystal x-ray solutions for \super. The literature composition corresponds to 10\% \textit{Ln}, 33\% Ti, and 57\% Sb. These results should be readily differentiable by standard EDS. 

Anecdotally, split site disorder can often be indicative of a misidentified unit cell. This could be consistent with our results, which can generally be represented as a 2$\times$2$\times$1 supercell. To investigate further, Figure \ref{fig:super}(b) shows several precession images generated from single-crystal scans on \NDsuper\ at the Cornell High Energy Synchrotron Source (CHESS). The cell is indexed in the \textit{Cmcm} (\textit{a,b,c}):(8.9,20.9,20.6\AA) cell. The top image shows a cut of the (HK0) plane, where we proceed to sum the intensities for each row ($\sum_k I_{hk0}$) to create a projection  of the H reflections on the right hand side. Dashed gray lines have been added to highlight all integer reflections. Similar results have been provided for the (HK0) and (0KL) plane. Our results indicate that all integer reflections are present, confirming the need for the larger \textit{a,b,c}):(8.9,20.9,20.6\AA) \textit{Cmcm} unit cell. 

Figure \ref{fig:super}(c) helps visualize two main differences that we want to highlight. First, the Ti site highlighted in purple for \super\ replaces the 25\% occupied, split Sb site in \disord. Correspondingly, you can see how the overall structure has adjusted to accommodate the Ti atom (the chains and neighboring Sb10 and Sb11 atoms are pushed away from the filled Ti-site. Second, the degree of splitting on the Sb10 and Sb11 sites is substantially reduced from 50/50\% in the literature result to approximately 90/10\% in our supercell. The degree of splitting in our refinement trends well with the rare-earth Shannon radius, increasing nearly linearly from 96/4\% with La to 87/13\% with Nd. The increased degree of splitting prefaces the destabilization of the phase with decreasing rare-earth radius, as compounds utilizing Sm were not able to be realized in this study. In fact, the competition between \super\ and the competing phases with smaller rare-earth ions will be discussed in a future manuscript. Some of the remaining Sb10/Sb11 splitting could be related to the thermal profile during growth, as the aggressive quenching (950\degree C) required to maintain clean crystals may contribute to the disorder. 

Ultimately our work suggests that \super\ may be the parent, ordered phase of \disord.\cite{bie2007ternary} Considering the different compositions, it seems unlikely that the two phases could be confused. Accordingly, we stress that we are not claiming that the previous work was incorrectly identified. We cannot exclude the possibility of a Ti--Sb solid-solution between \super\ and \disord\ that could account for the disorder and compositional differences between the two phases. Additionally, given our observations of the temperature-dependent instability of \super\ in the Sn flux, there may be more complex thermodynamic considerations at play as well. Finally, despite our polycrystalline results identifying that the pure antimonide exists, it is possible that a small but undetectable percentage of Sn (incorporated from the flux) may stabilize the longer-range order in \super.

\subsection{Electronic Structure}

\begin{figure}
\includegraphics[width=\linewidth]{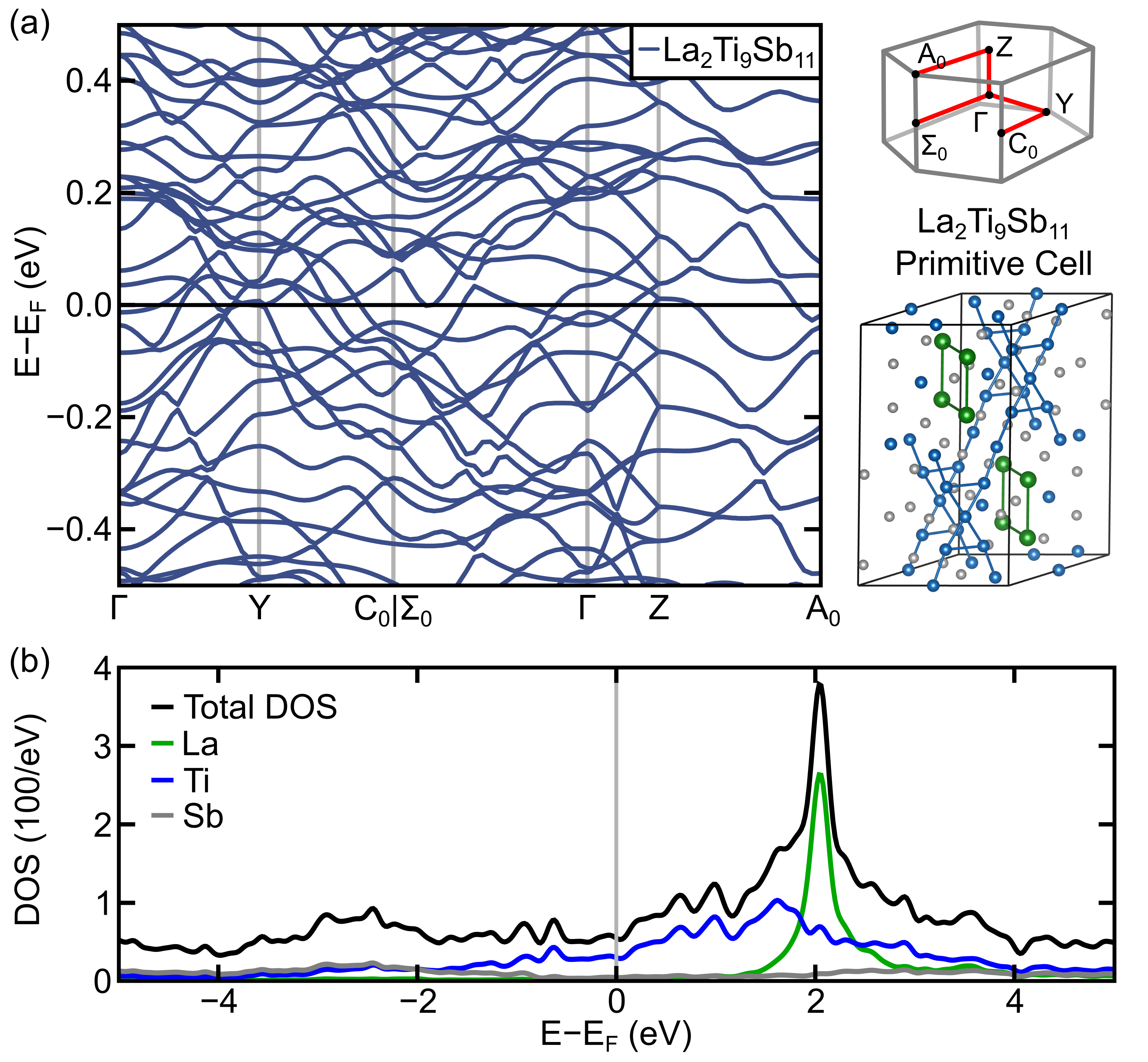}
\caption{(a) The electronic structure of the \super family is quite complex, owing to the large, low symmetry (\textit{Cmcm}) unit cell. To simplify the analysis, we show only calculations on the nonmagnetic La$_2$Ti$_9$Sb$_{11}$ compound and highlight only results within $\pm$0.5~eV around the Fermi level. (b) DOS calculations highlight that the states near E$_\text{F}$ are dominated by Ti orbitals, a feature shared amongst many Ti--Ti network based compounds like the \textit{Ln}Ti$_3$Bi$_4$ and \textit{Ln}$_{2-x}$Ti$_{6+x}$Bi$_9$ metals.\cite{ortiz2024intricate}}
\label{fig:dft}
\end{figure}

\begin{figure*}
\includegraphics[width=1\textwidth]{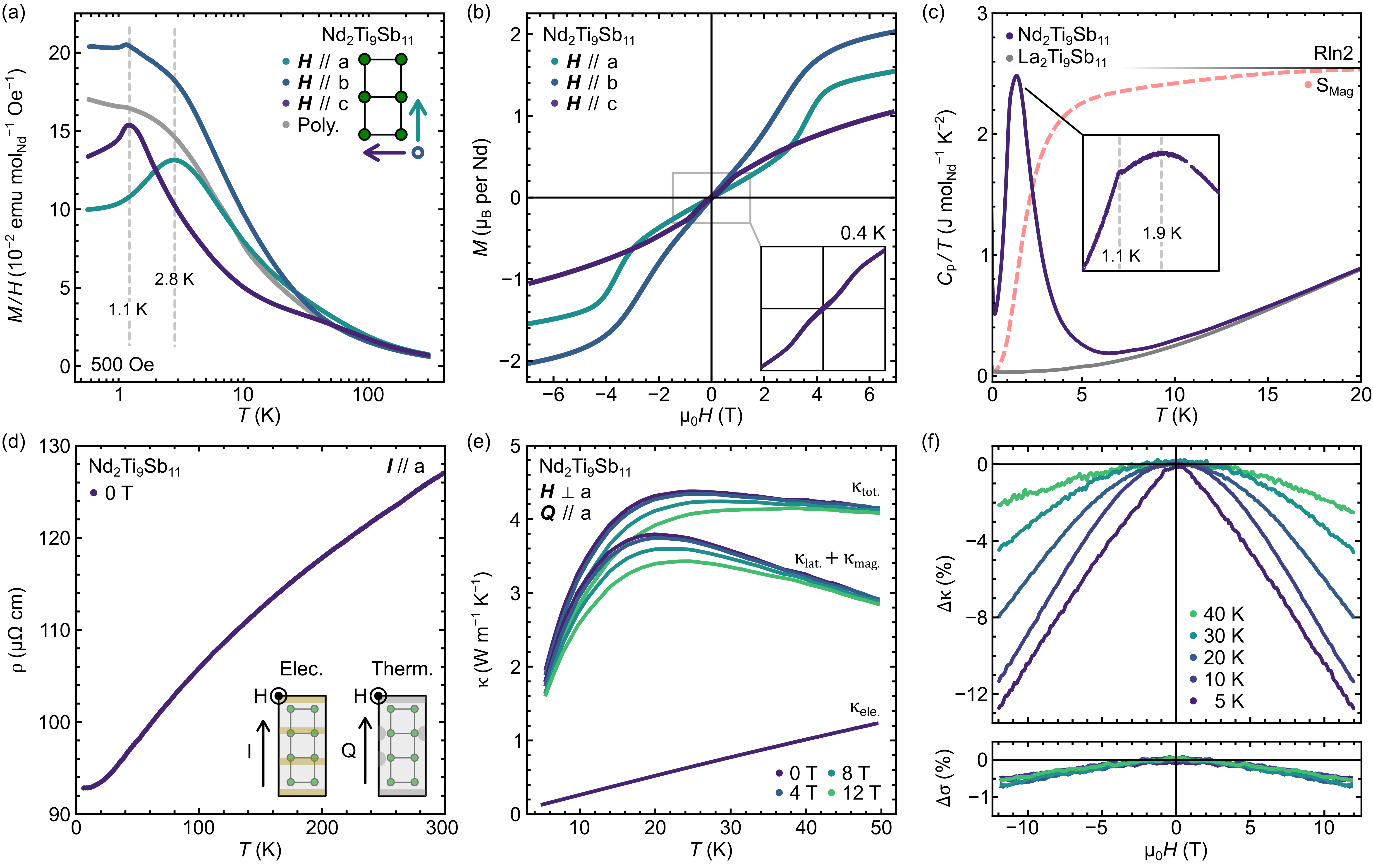}
\caption{Temperature-dependent magnetization (a) highlights the orientation-specific response along the three principle crystallographic directions in \NDsuper. Multiple temperature scales are noted, approximately 1.1~K for the cusp/plateau along \textit{b} and \textit{c}, and a broad peak centered around 2.8~K along \textit{a}. The inset shows these orientations relative to the spin ladder orientation. Isothermal magnetization (b) presents a similar set of orientation-dependent measurements, highlighting metamagnetic-like transitions at high fields. Heat capacity measurements (c) identify two temperature scales, a weak sharper feature at 1.1~K and a broad feature centered around 1.9~K. The entropy release by approximately 5~K approaches the $R\ln2$ expected of a ground state doublet. Electronic resistivity (d) confirms metallic nature of the system and provides an estimate of the electronic thermal conductivity (e) used to determine the lattice and potential magnon thermal transport. The thermal conductivity of \NDsuper\ is strongly suppressed by applied field (f), while the electronic transport is nearly field-independent (f, lower panel).}
\label{fig:Nd}
\end{figure*}

Here we turn to briefly examine the electronic structure of \super. As one may expect from the large orthorhombic cell, the electronic structure (Figure \ref{fig:dft}) is complex. To eliminate the need for magnetic simulations in the already complex calculation, we have chosen to investigate only the nonmagnetic \LAsuper~ compound. Similarly, we have chosen to ignore spin-orbit effects, producing a basic foundation for understanding the electronic structure. 

Figure \ref{fig:dft} shows the resulting electronic structure in a small window around the Fermi level ($E_\text{F}\pm0.5$eV). The simplified \LAsuper~ primitive cell and the calculated high-symmetry path are shown to the right of the band structure. Additional density-of-states (DOS) calculations are shown below the dispersion. Band structure calculations suggest that \super\ compounds are metallic, with the majority of states at near the Fermi level dominated by Ti orbitals. This is similar to previous results in systems with Ti--Ti networks like the Ti-based kagome materials \textit{Ln}Ti$_3$Sb$_4$ and shurikagome \textit{Ln}$_{2-x}$Ti$_{6+x}$Bi$_9$.\cite{ortiz2024intricate}

The metallic nature of the \super\ family may be pertinent to the magnetic properties, as extended interactions between \textit{Ln} atoms is likely mediated \textit{via} the Ruderman-Kittel-Kasuya-Yosida (RKKY) interaction. The similar \textit{Ln}--\textit{Ln} leg and rung separations (4.4--4.6\AA) suggests comparable magnetic interactions (e.g. J$_\text{leg}$ $\sim$ J$_\text{rung}$). These interactions are likely much stronger ($>$5-10) than the inter-ladder interactions. The nearest ladder center-to-center distance is approximately 10.5\AA~, though the closest next-nearest neighbor magnetic interaction is 7.8\AA~ (interaction between right leg to left leg of nearest diagonal ladder). Regardless, the rung/leg interactions within each ladder are expected to dominate. It is possible that secondary effects (e.g. Fermi surface anisotropy) can lead to different magnitudes of J$_\text{leg}$ and J$_\text{rung}$, but with the complex electronic structure, it is not trivial to tell.

\subsection{Physical properties of \NDsuper}

We now turn to examine the basic physical properties of the \super\ (\textit{Ln}:La--Nd) compounds. We focus on the Nd-based compound as the transition temperatures are the highest and the features are the strongest. We will briefly discuss the properties of the Ce- and Pr-based compounds afterwards, with potential extensions towards other systems at the very end. 

Figure \ref{fig:Nd} presents a series of basic physical properties measurements on single crystals of \NDsuper. The crystal structure is orthorhombic, so we need to examine three directions at a minimum. Figure \ref{fig:Nd}(a) highlights three magnetization measurements for \NDsuper\ with a 500~Oe field oriented along each of the three principle axes. We have also included a polycrystalline measurement on a ground single crystal for reference. The inset of Figure \ref{fig:Nd}(a) provides a quick reference for the orientation of the field relative to the \textit{Ln} spin-ladders. In the case of Nd, the system shows a broad hump around 3~K when $H\parallel a$, a sharper cusp at 1.1~K when $H\parallel c$, and a plateau when $H\parallel b$. 

Curie-Weiss analysis of the polycrystalline data is challenging, as the inverse susceptibility data contains three linear regimes (50--300~K, 20--40~K, and 5--10~K), with respective $\mu_\text{eff}$ of 4.3$\mu_\text{B}$, 3.6$\mu_\text{B}$ and 3.1$\mu_\text{B}$ (see ESI\cite{ESI}). In general, the likelihood of low-lying crystal field levels will complicate our analysis. Occasionally, fits to the low-temperature data can be helpful in avoiding crystal-field effects as the levels are thermally populated, but choosing the temperature range is purely speculative without more detailed crystal field measurements. This is also complicated by the low dimensionality of the system, which can exacerbate magnetic fluctuations in the paramagnetic state. As such, we will omitted further discussion of the Curie-Weiss fits. For completeness, we have included inverse susceptibility measurements and the Curie-Weiss parameters for the high-temperature fits in the ESI.\cite{ESI}

Figure \ref{fig:Nd}(b) presents the isothermal magnetization as a function of magnetic field for the three orientations described in Figure \ref{fig:Nd}(a). All of the orientations examined here exhibit features that are reminiscent of spin-reorientation effects, with H$\parallel$a exhibiting data most similar to a spin-flop. Some subtle coercivity and changes in curvature are noted for the H$\parallel$c orientation (inset) at low fields. Taken together with the cusp-like features from Figure \ref{fig:Nd}(a), \NDsuper\ appears to exhibit antiferromagnetic interactions within the Nd spin-ladder sublattice. We note that none of the curves approach the expected $gJ$ for a Nd$^{3+}$ free-ion ($\sim$ 3.27$\mu_\text{B}$) by 7~T. However, this is perhaps expected considering the quasi-1D nature of the spin-ladder and the complex chemical environment surrounding the Nd$^{3+}$. Similar divergences from the free-ion $gJ$ have been observed in other spin-ladder and spin-chain systems.\cite{zeng2024k2renb5o15}

The various temperature scales observed in Figure \ref{fig:Nd}(a) complicate our supposition of an ordered antiferromagnetic ground state based on the magnetization alone. The broad feature observed around 2.8~K, for example, is reminiscent of the build-up of magnetic correlations often observed alongside low-dimensional magnetism. As such, we turn to the heat-capacity (Figure \ref{fig:Nd}(c)) to help clarify. We present the $C_\text{p}/T$ data for \NDsuper using the non-magnetic analog \LAsuper~ as a phonon reference. The integrated magnetic entropy is shown as the pink dashed line.

We first note that the entropy release is quite broad, persisting up to 5-10~K. Within the primary heat capacity peak there are two features: 1) a broad peak centered around 1.9~K in $C_\text{p}$, and 2) a weak, sharper feature at 1.1~K. The weak feature corresponds with the cusp observed in both the H$\parallel$b and H$\parallel$c directions, while the broad feature vaguely corresponds to the downturn in the broad hump seen most strongly when H$\parallel$a. The net entropy release by $\sim$10~K is about 90\% that of the expected $R\ln2$ for a magnetic doublet ground state. Qualitatively similar heat capacity results have been observed in other low-dimensional magnetic systems like $S=1/2$ Cu in Cu(C$_5$H$_5$NO)$_6$(BF$_4$)$_2$, where the sharper feature is proposed as the transition to a magnetically ordered ground state.\cite{algra1978one} We note that the nonmagnetic subtraction is less accurate at temperatures above 50~K, and so we limit our analysis to the low-temperature regime.

In magnetic spin ladder compounds, one can often observe a rich variety of interesting anisotropic electrical and thermal transport properties due to their reduced dimensionality and diverse magnetic ground states.\cite{sologubenko2007thermal,sologubenko2001heat,chernyshev2005thermal,klumper2002thermal,sologubenko2000thermal} The most intriguing transport phenomena are typically observed when the electrical or heat current flows along the spin chain or ladder direction. We thus studied the temperature and field dependence of electrical and thermal transport properties of \NDsuper. For convenience, the inset in the lower-right hand of Figure \ref{fig:Nd}(d) shows the sample geometry for the following electronic and thermal transport experiments. Figure \ref{fig:Nd}(d) establishes that \NDsuper\ is a poor metal with a high residual resistivity $\sim$0.1m$\Omega$-cm and a low residual resistivity ratio (RRR) of 1.3-1.4 (e.g. the resistivity is largely temperature-independent). 

\begin{figure*}
\includegraphics[width=1\textwidth]{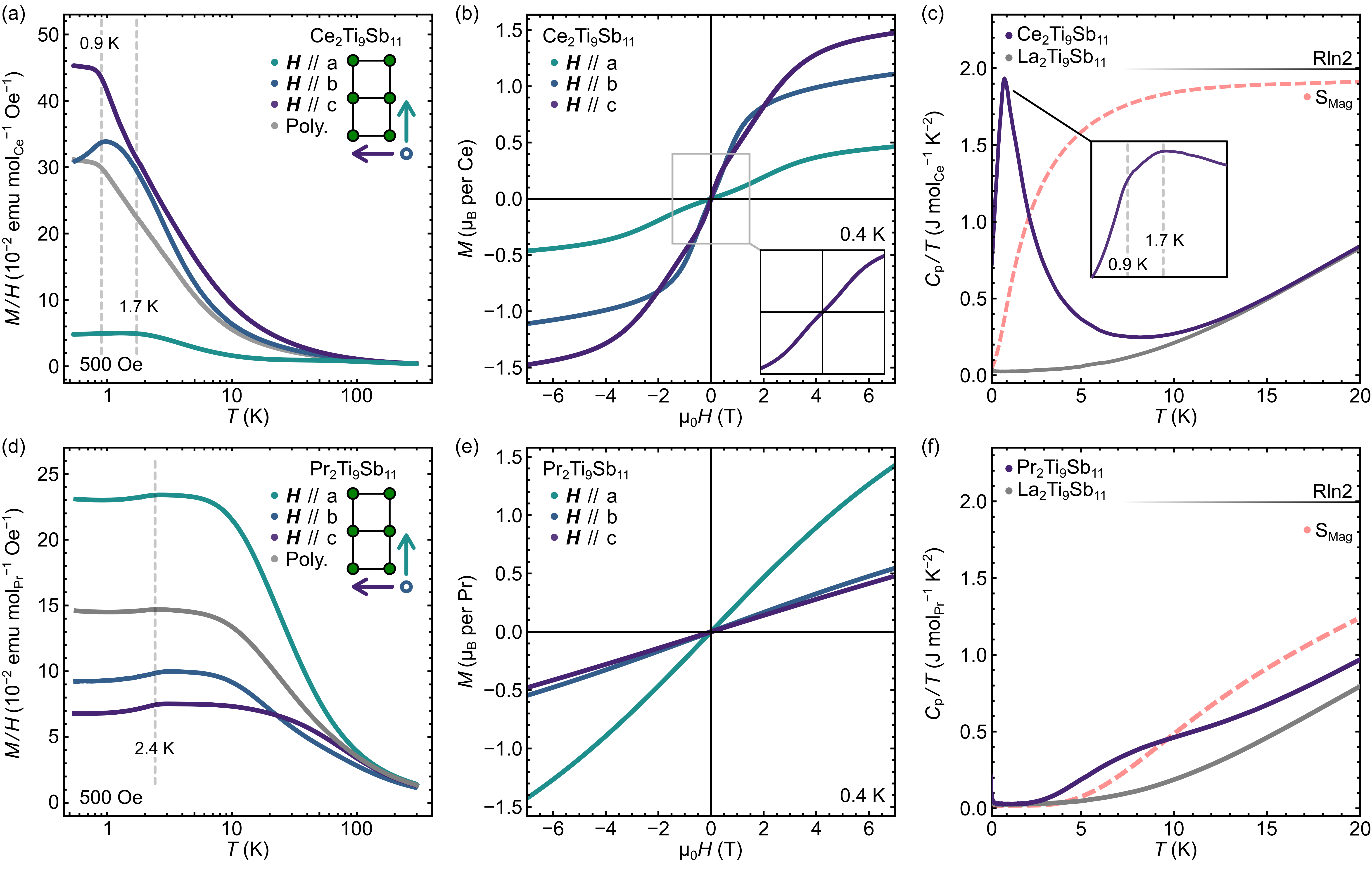}
\caption{Here we provide a brief assessment of the magnetic properties of \CEsuper\ and \PRsuper. The temperature-dependent magnetization of \CEsuper\ (a) is reminiscent of \NDsuper, showing a plateau, a cusp, and a broad peak. The isothermal magnetization (b) does not show as pronounced metamagnetic-like features but the curves still exhibit field-dependent changes. The heat capacity for \CEsuper\ shows two features, a broad peak centered around 1.7~K and a shoulder at 0.9~K. These correspond well to the temperature scales seen in the magnetization. The magnetic entropy release approaches $R\ln2$ by approximately 5-10~K, consistent with a magnetic doublet ground state. The temperature-dependent magnetization of \PRsuper\ (d) is largely featureless, with only a broad saturation near 10-20~K. The isothermal magnetization (e) is similarly featureless and non saturating. Both results are consistent with the heat capacity (f) which only exhibits a broad release of magnetic entropy over an extended 2-50~K temperature range. These results are consistent with a non-magnetic singlet state for the non-Kramers Pr$^{3+}$.
}
\label{fig:CePr}
\end{figure*}

Figure \ref{fig:Nd}(e) shows the temperature- and field-dependent thermal conductivity with the heat flow directed parallel to the spin-ladder (\textit{a}-axis) and magnetic fields applied perpendicular to the \textit{a}-axis (H $\parallel \textit{b}$). We subsequently estimate the electronic thermal conductivity using the Wiedemann-Franz law and the measured electronic conductivity (resistivity) shown in Figure \ref{fig:Nd}(d). The electronic transport is nearly field-independent up to 12~T (see lower panel of Figure \ref{fig:Nd}(f)). As such, $\kappa_\text{ele.}$ is effectively identical for all fields, allowing us to easily isolate the lattice and potential magnon contributions to the thermal conductivity $\kappa_\text{lat.}+\kappa_\text{mag.}$ from $\kappa_\text{tot.}$. Considering the low energy scale of rare-earth magnetism, we measured the thermal conductivity up to 50\.K. The data is relatively typical of phonon heat transport, with a peak centered around 20\.K signaling the high quality of the crystals. 

Figure \ref{fig:Nd}(f) summarizes the percent change in the thermal conductivity (upper panel) and the electrical conductivity (lower panel) as a function of magnetic field. The effect on the electrical transport is markedly weak, with less than a 1\% change under a 12~T field. This contrasts sharply with the thermal conductivity of \NDsuper, which is strongly suppressed by an applied magnetic field. All field-dependent thermal transport data were collected in the paramagnetic state within the temperature range 5-40\,K. At 5\,K, a 12\% suppression is observed  at 12\,T. This suppression remains observable even at 40\,K, 40 times larger than T$_N$. We also examined the case when H is parallel to the spin-ladder and observed qualitatively similar results. 

Given the negligible field dependence of the electrical conductivity, the observed suppression of thermal conductivity by magnetic fields is likely due to the enhanced phonon scattering from magnetic fluctuations. We note that the suppression of the thermal conductivity also appears to be strongest near the thermal conductivity peak around 20~K. This response could be attributed to something more exotic like phonon-drag, and certainly suggests some degree of strong spin-lattice coupling in \NDsuper. However, additional measurements (e.g. Seebeck coefficient, neutron scattering, additional transport results) would be needed to investigate further. The observed field-induced suppression of thermal conductivity is also surprising because external magnetic fields typically align the magnetic moments and reduce magnetic fluctuations in the paramagnetic state. This unusual behavior may stem from the unique spin ladder structure of \NDsuper\ and further insight into potential spin-lattice coupling could be obtained through inelastic neutron scattering studies, dilatometry, or thermal expansion experiments. 

\subsection{Physical properties of \CEsuper\ and \PRsuper}

The slightly higher transition temperatures in \NDsuper\ make the analysis of the magnetic and thermodynamic properties somewhat easier to tackle. However, the structurally analogous \CEsuper\ and \PRsuper\ both exemplify how the choice of rare-earth can be used to tune the magnetic anisotropy, strength, and nature of the ground state in rare-earth intermetallics. We will only present a brief discussion of these two materials, at least in comparison to \NDsuper, leaving more detailed studies for future efforts. 

Figure \ref{fig:CePr}(a-c) present the basic magnetization and heat capacity results for a single crystal of \CEsuper. Figure \ref{fig:CePr}(a) demonstrates the anisotropic magnetic response along the three principle axes of the crystal. Some directions are qualitatively similar to those observed in \NDsuper. H$\parallel$b resembles the broad hump or cusp, and H$\parallel$c is reminiscent of the plateau. However, the magnetic response along H$\parallel$a in \CEsuper\ is extremely weak, suggesting that the moments appear to lie predominately in the b-c plane. 

The isothermal magnetization in Figure \ref{fig:CePr}(b) is reminiscent of the data from \NDsuper\ as well, with each direction exhibiting markedly non-linear modulations in the magnetization with field. The effect is less pronounced compared to the metamagnetic-like features seen in \NDsuper, perhaps due to the lower transition temperatures observed in \CEsuper. Isothermal magnetization only recovers 70\% of the expected $gJ=2.1\mu_\text{B}$ for Ce$^{3+}$ when H$\parallel$c.

Figure \ref{fig:CePr}(c) highlights the low-temperature $C_\text{p}/T$ for \CEsuper, with the La-based non-magnetic analog shown for comparison. As observed in \NDsuper, the transition is broad and contains two temperatures scales: 1) a hump that peaks in $C_\text{P}$ around 1.7~K, and 2) a secondary (weak) feature that occurs around 0.9~K. These temperatures correlate well with the broad rise in H$\parallel$a and the plateau/cusp observed in H$\parallel$b and H$\parallel$c. 

The magnetization and heat capacity results in \CEsuper\ are vaguely reminiscent of the results for \NDsuper. Both materials present magnetization results containing: 1) a plateau, 2) a cusp near the plateau temperature, and 3) a higher-temperature broad hump. \CEsuper\ shows a substantially weaker response in the isothermal magnetization (Figure \ref{fig:CePr}(b)). We suspect that \CEsuper\ exhibits antiferromagnetic interactions similar to \NDsuper, though with a lower temperature scale perhaps suggestive of Ce$^{3+}$'s weaker $J_\text{eff}=1/2$ moment.

The case of \PRsuper\ is less clear. Owing to the non-Kramers nature of the Pr$^{3+}$ cation, Pr-based intermetallics can be particularly sensitive to disorder, complex chemical environments, and crystal field effects. In many cases, this causes Pr to be weakly magnetic, adopting a ``non-magnetic'' singlet ground state. Figure \ref{fig:CePr}(d-f) summarize the basic magnetization and heat capacity measurements on \PRsuper. 

Magnetization measurements (Figure \ref{fig:CePr}(d)) are broad and largely featureless. There is a weak downturn that occurs near 2.6--3.0~K in all field orientations, though this feature does not have a corresponding feature in the heat capacity. The isothermal magnetization measurements (Figure \ref{fig:CePr}(e)) are non-saturating and featureless, consistent with a non-magnetic singlet ground state. Finally, the heat capacity (Figure \ref{fig:CePr}(f)) only features an extended release of magnetic entropy which begins well above 20~K. The broad peak-like feature centered around 5-10~K, corresponding to the saturation of the magnetization measurements in Figure \ref{fig:CePr}(d), recovers only a small fraction of the entropy expected for a doublet ground state, corroborating our suspicion of a non-magnetic singlet state. 

Despite the non-magnetic ground state for \PRsuper, the complex low-dimensional antiferromagnetism in \CEsuper and \NDsuper offers a unique material platform for the exploration of two-legged spin ladders. A summary of some key magnetic, thermodynamic, and crystallographic parameters can be found in the ESI.\cite{ESI} While it appears that Nd is the smallest rare-earth that currently stabilizes the structure, perhaps chemical modifications to the supporting lattices (e.g. substitution of Sb with As) could realize new rare-earth variants of the structure. Besides the potentially higher critical temperatures as we move towards Gd/Tb, the ability to further tune the potential anisotropies and effective $J$ of the spin-ladder offers a unique route to new model systems.

\section{Conclusions}

We began this work by presenting the discovery of a new family of rare-earth spin ladder compounds \super\ (\textit{Ln}: La--Nd). The crystal structure of the \super\ family is complex, but hosts well-isolated, square spin ladders. The ladders are interwoven with a complex Ti--Ti and Ti--Sb network which carries traces of the Ti-based kagome lattice seen in the \textit{Ln}Ti$_3$Bi$_4$ phases. We discussed potential structural relations to the \disord\ family, which possess qualitatively similar features but non-trivial differences in the chemical stoichometry and substantial disorder on the Sb sublattice. The \NDsuper\ and \CEsuper\ compounds exhibit complex low-temperature magnetism consistent with a low-dimensional antiferromagnetic ground state. Additional electronic and thermal transport measurements on \NDsuper\ reveal an unexpected and pronounced suppression of the thermal conductivity with applied magnetic field in the paramagnetic state. Together our results present a new spin-ladder compound utilizing the tunable magnetism of the rare-earth elements. We hope that further chemical modifications may help stabilize smaller rare-earth variants, presenting a new route to explore spin ladder physics.

\section{Acknowledgements}

This work was supported by the U.S. Department of Energy (DOE), Office of Science, Basic Energy Sciences (BES), Materials Sciences and Engineering Division. This research used resources of the Compute and Data Environment for Science (CADES) at the Oak Ridge National Laboratory, which is supported by the Office of Science of the U.S. Department of Energy under Contract No. DE-AC05-00OR22725. Research conducted at the Center for High-Energy X-ray Sciences (CHEXS) is supported by the NSF (BIO, ENG, and MPS Directorates) under award DMR-1829070. We thank Prof. Ramesh Nath for his helpful discussions.\footnote{Notice: This manuscript has been authored by UT-Battelle, LLC, under contract DE-AC05-00OR22725 with the US Department of Energy (DOE). The US government retains and the publisher, by accepting the article for publication, acknowledges that the US government retains a nonexclusive, paid-up, irrevocable, worldwide license to publish or reproduce the published form of this manuscript, or allow others to do so, for US government purposes. DOE will provide public access to these results of federally sponsored research in accordance with the DOE Public Access Plan (https://www.energy.gov/doe-public-access-plan).} 

\bibliography{LnTi9Sb11_Ladders}

\providecommand{\noopsort}[1]{}\providecommand{\singleletter}[1]{#1}%
\begin{thebibliography}{54}%
\makeatletter
\providecommand \@ifxundefined [1]{%
 \@ifx{#1\undefined}
}%
\providecommand \@ifnum [1]{%
 \ifnum #1\expandafter \@firstoftwo
 \else \expandafter \@secondoftwo
 \fi
}%
\providecommand \@ifx [1]{%
 \ifx #1\expandafter \@firstoftwo
 \else \expandafter \@secondoftwo
 \fi
}%
\providecommand \natexlab [1]{#1}%
\providecommand \enquote  [1]{``#1''}%
\providecommand \bibnamefont  [1]{#1}%
\providecommand \bibfnamefont [1]{#1}%
\providecommand \citenamefont [1]{#1}%
\providecommand \href@noop [0]{\@secondoftwo}%
\providecommand \href [0]{\begingroup \@sanitize@url \@href}%
\providecommand \@href[1]{\@@startlink{#1}\@@href}%
\providecommand \@@href[1]{\endgroup#1\@@endlink}%
\providecommand \@sanitize@url [0]{\catcode `\\12\catcode `\$12\catcode
  `\&12\catcode `\#12\catcode `\^12\catcode `\_12\catcode `\%12\relax}%
\providecommand \@@startlink[1]{}%
\providecommand \@@endlink[0]{}%
\providecommand \url  [0]{\begingroup\@sanitize@url \@url }%
\providecommand \@url [1]{\endgroup\@href {#1}{\urlprefix }}%
\providecommand \urlprefix  [0]{URL }%
\providecommand \Eprint [0]{\href }%
\providecommand \doibase [0]{https://doi.org/}%
\providecommand \selectlanguage [0]{\@gobble}%
\providecommand \bibinfo  [0]{\@secondoftwo}%
\providecommand \bibfield  [0]{\@secondoftwo}%
\providecommand \translation [1]{[#1]}%
\providecommand \BibitemOpen [0]{}%
\providecommand \bibitemStop [0]{}%
\providecommand \bibitemNoStop [0]{.\EOS\space}%
\providecommand \EOS [0]{\spacefactor3000\relax}%
\providecommand \BibitemShut  [1]{\csname bibitem#1\endcsname}%
\let\auto@bib@innerbib\@empty
\bibitem [{\citenamefont {Haldane}(1983)}]{haldane1983nonlinear}%
  \BibitemOpen
  \bibfield  {author} {\bibinfo {author} {\bibfnamefont {F.~D.~M.}\
  \bibnamefont {Haldane}},\ }\bibfield  {title} {\bibinfo {title} {Nonlinear
  field theory of large-spin heisenberg antiferromagnets: semiclassically
  quantized solitons of the one-dimensional easy-axis n{\'e}el state},\
  }\href@noop {} {\bibfield  {journal} {\bibinfo  {journal} {Physical review
  letters}\ }\textbf {\bibinfo {volume} {50}},\ \bibinfo {pages} {1153}
  (\bibinfo {year} {1983})}\BibitemShut {NoStop}%
\bibitem [{\citenamefont {Bethe}(1931)}]{bethe1931theorie}%
  \BibitemOpen
  \bibfield  {author} {\bibinfo {author} {\bibfnamefont {H.}~\bibnamefont
  {Bethe}},\ }\bibfield  {title} {\bibinfo {title} {Zur theorie der metalle: I.
  eigenwerte und eigenfunktionen der linearen atomkette},\ }\href@noop {}
  {\bibfield  {journal} {\bibinfo  {journal} {Zeitschrift f{\"u}r Physik}\
  }\textbf {\bibinfo {volume} {71}},\ \bibinfo {pages} {205} (\bibinfo {year}
  {1931})}\BibitemShut {NoStop}%
\bibitem [{\citenamefont {Hulth{\'e}n}(1938)}]{hulthen1938austauschproblem}%
  \BibitemOpen
  \bibfield  {author} {\bibinfo {author} {\bibfnamefont {L.}~\bibnamefont
  {Hulth{\'e}n}},\ }\emph {\bibinfo {title} {{\"U}ber das austauschproblem
  eines kristalles}},\ \href@noop {} {Ph.D. thesis},\ \bibinfo  {school}
  {Almqvist \& Wiksell} (\bibinfo {year} {1938})\BibitemShut {NoStop}%
\bibitem [{\citenamefont {Sutherland}(1970)}]{sutherland1970two}%
  \BibitemOpen
  \bibfield  {author} {\bibinfo {author} {\bibfnamefont {B.}~\bibnamefont
  {Sutherland}},\ }\bibfield  {title} {\bibinfo {title} {Two-dimensional
  hydrogen bonded crystals without the ice rule},\ }\href@noop {} {\bibfield
  {journal} {\bibinfo  {journal} {Journal of Mathematical Physics}\ }\textbf
  {\bibinfo {volume} {11}},\ \bibinfo {pages} {3183} (\bibinfo {year}
  {1970})}\BibitemShut {NoStop}%
\bibitem [{\citenamefont {Baxter}(1971)}]{baxter1971one}%
  \BibitemOpen
  \bibfield  {author} {\bibinfo {author} {\bibfnamefont {R.}~\bibnamefont
  {Baxter}},\ }\bibfield  {title} {\bibinfo {title} {One-dimensional
  anisotropic heisenberg chain},\ }\href@noop {} {\bibfield  {journal}
  {\bibinfo  {journal} {Physical Review Letters}\ }\textbf {\bibinfo {volume}
  {26}},\ \bibinfo {pages} {834} (\bibinfo {year} {1971})}\BibitemShut
  {NoStop}%
\bibitem [{\citenamefont {Johnson}\ \emph {et~al.}(1973)\citenamefont
  {Johnson}, \citenamefont {Krinsky},\ and\ \citenamefont
  {McCoy}}]{johnson1973vertical}%
  \BibitemOpen
  \bibfield  {author} {\bibinfo {author} {\bibfnamefont {J.~D.}\ \bibnamefont
  {Johnson}}, \bibinfo {author} {\bibfnamefont {S.}~\bibnamefont {Krinsky}},\
  and\ \bibinfo {author} {\bibfnamefont {B.~M.}\ \bibnamefont {McCoy}},\
  }\bibfield  {title} {\bibinfo {title} {Vertical-arrow correlation length in
  the eight-vertex model and the low-lying excitations of the x- y- z
  hamiltonian},\ }\href@noop {} {\bibfield  {journal} {\bibinfo  {journal}
  {Physical Review A}\ }\textbf {\bibinfo {volume} {8}},\ \bibinfo {pages}
  {2526} (\bibinfo {year} {1973})}\BibitemShut {NoStop}%
\bibitem [{\citenamefont {Kl{\"u}mper}(1993)}]{klumper1993thermodynamics}%
  \BibitemOpen
  \bibfield  {author} {\bibinfo {author} {\bibfnamefont {A.}~\bibnamefont
  {Kl{\"u}mper}},\ }\bibfield  {title} {\bibinfo {title} {Thermodynamics of the
  anisotropic spin-1/2 heisenberg chain and related quantum chains},\
  }\href@noop {} {\bibfield  {journal} {\bibinfo  {journal} {Zeitschrift
  f{\"u}r Physik B Condensed Matter}\ }\textbf {\bibinfo {volume} {91}},\
  \bibinfo {pages} {507} (\bibinfo {year} {1993})}\BibitemShut {NoStop}%
\bibitem [{\citenamefont {Anderson}(1952)}]{anderson1952approximate}%
  \BibitemOpen
  \bibfield  {author} {\bibinfo {author} {\bibfnamefont {P.~W.}\ \bibnamefont
  {Anderson}},\ }\bibfield  {title} {\bibinfo {title} {An approximate quantum
  theory of the antiferromagnetic ground state},\ }\href@noop {} {\bibfield
  {journal} {\bibinfo  {journal} {Physical Review}\ }\textbf {\bibinfo {volume}
  {86}},\ \bibinfo {pages} {694} (\bibinfo {year} {1952})}\BibitemShut
  {NoStop}%
\bibitem [{\citenamefont {Anderson}(1951)}]{anderson1951limits}%
  \BibitemOpen
  \bibfield  {author} {\bibinfo {author} {\bibfnamefont {P.}~\bibnamefont
  {Anderson}},\ }\bibfield  {title} {\bibinfo {title} {Limits on the energy of
  the antiferromagnetic ground state},\ }\href@noop {} {\bibfield  {journal}
  {\bibinfo  {journal} {Physical Review}\ }\textbf {\bibinfo {volume} {83}},\
  \bibinfo {pages} {1260} (\bibinfo {year} {1951})}\BibitemShut {NoStop}%
\bibitem [{\citenamefont {Kubo}(1952)}]{kubo1952spin}%
  \BibitemOpen
  \bibfield  {author} {\bibinfo {author} {\bibfnamefont {R.}~\bibnamefont
  {Kubo}},\ }\bibfield  {title} {\bibinfo {title} {The spin-wave theory of
  antiferromagnetics},\ }\href@noop {} {\bibfield  {journal} {\bibinfo
  {journal} {Physical Review}\ }\textbf {\bibinfo {volume} {87}},\ \bibinfo
  {pages} {568} (\bibinfo {year} {1952})}\BibitemShut {NoStop}%
\bibitem [{\citenamefont {Dagotto}\ and\ \citenamefont
  {Rice}(1996)}]{dagotto1996surprises}%
  \BibitemOpen
  \bibfield  {author} {\bibinfo {author} {\bibfnamefont {E.}~\bibnamefont
  {Dagotto}}\ and\ \bibinfo {author} {\bibfnamefont {T.}~\bibnamefont {Rice}},\
  }\bibfield  {title} {\bibinfo {title} {Surprises on the way from one-to
  two-dimensional quantum magnets: The ladder materials},\ }\href@noop {}
  {\bibfield  {journal} {\bibinfo  {journal} {Science}\ }\textbf {\bibinfo
  {volume} {271}},\ \bibinfo {pages} {618} (\bibinfo {year}
  {1996})}\BibitemShut {NoStop}%
\bibitem [{\citenamefont {Dagotto}(1999)}]{dagotto1999experiments}%
  \BibitemOpen
  \bibfield  {author} {\bibinfo {author} {\bibfnamefont {E.}~\bibnamefont
  {Dagotto}},\ }\bibfield  {title} {\bibinfo {title} {Experiments on ladders
  reveal a complex interplay between a spin-gapped normal state and
  superconductivity},\ }\href@noop {} {\bibfield  {journal} {\bibinfo
  {journal} {Reports on Progress in Physics}\ }\textbf {\bibinfo {volume}
  {62}},\ \bibinfo {pages} {1525} (\bibinfo {year} {1999})}\BibitemShut
  {NoStop}%
\bibitem [{\citenamefont {Kageyama}\ \emph {et~al.}(2008)\citenamefont
  {Kageyama}, \citenamefont {Watanabe}, \citenamefont {Tsujimoto},
  \citenamefont {Kitada}, \citenamefont {Sumida}, \citenamefont {Kanamori},
  \citenamefont {Yoshimura}, \citenamefont {Hayashi}, \citenamefont {Muranaka},
  \citenamefont {Takano} \emph {et~al.}}]{kageyama2008spin}%
  \BibitemOpen
  \bibfield  {author} {\bibinfo {author} {\bibfnamefont {H.}~\bibnamefont
  {Kageyama}}, \bibinfo {author} {\bibfnamefont {T.}~\bibnamefont {Watanabe}},
  \bibinfo {author} {\bibfnamefont {Y.}~\bibnamefont {Tsujimoto}}, \bibinfo
  {author} {\bibfnamefont {A.}~\bibnamefont {Kitada}}, \bibinfo {author}
  {\bibfnamefont {Y.}~\bibnamefont {Sumida}}, \bibinfo {author} {\bibfnamefont
  {K.}~\bibnamefont {Kanamori}}, \bibinfo {author} {\bibfnamefont
  {K.}~\bibnamefont {Yoshimura}}, \bibinfo {author} {\bibfnamefont
  {N.}~\bibnamefont {Hayashi}}, \bibinfo {author} {\bibfnamefont
  {S.}~\bibnamefont {Muranaka}}, \bibinfo {author} {\bibfnamefont
  {M.}~\bibnamefont {Takano}}, \emph {et~al.},\ }\bibfield  {title} {\bibinfo
  {title} {Spin-ladder iron oxide: Sr3fe2o5},\ }\href@noop {} {\bibfield
  {journal} {\bibinfo  {journal} {Angewandte Chemie}\ }\textbf {\bibinfo
  {volume} {120}},\ \bibinfo {pages} {5824} (\bibinfo {year}
  {2008})}\BibitemShut {NoStop}%
\bibitem [{\citenamefont {Hiroi}\ \emph {et~al.}(1991)\citenamefont {Hiroi},
  \citenamefont {Azuma}, \citenamefont {Takano},\ and\ \citenamefont
  {Bando}}]{hiroi1991new}%
  \BibitemOpen
  \bibfield  {author} {\bibinfo {author} {\bibfnamefont {Z.}~\bibnamefont
  {Hiroi}}, \bibinfo {author} {\bibfnamefont {M.}~\bibnamefont {Azuma}},
  \bibinfo {author} {\bibfnamefont {M.}~\bibnamefont {Takano}},\ and\ \bibinfo
  {author} {\bibfnamefont {Y.}~\bibnamefont {Bando}},\ }\bibfield  {title}
  {\bibinfo {title} {A new homologous series srn- 1cun+ 1o2n found in the
  sro-cuo system treated under high pressure},\ }\href@noop {} {\bibfield
  {journal} {\bibinfo  {journal} {Journal of Solid State Chemistry}\ }\textbf
  {\bibinfo {volume} {95}},\ \bibinfo {pages} {230} (\bibinfo {year}
  {1991})}\BibitemShut {NoStop}%
\bibitem [{\citenamefont {Koo}\ and\ \citenamefont
  {Whangbo}(1999)}]{koo1999analysis}%
  \BibitemOpen
  \bibfield  {author} {\bibinfo {author} {\bibfnamefont {H.-J.}\ \bibnamefont
  {Koo}}\ and\ \bibinfo {author} {\bibfnamefont {M.-H.}\ \bibnamefont
  {Whangbo}},\ }\bibfield  {title} {\bibinfo {title} {Analysis of the
  spin--spin interactions in layered oxides $\alpha$-nav2o5, cav2o5 and mgv2o5
  and the spin-peierls distortion in $\alpha$-nav2o5 by molecular orbital,
  madelung energy and bond valence sum calculations},\ }\href@noop {}
  {\bibfield  {journal} {\bibinfo  {journal} {Solid state communications}\
  }\textbf {\bibinfo {volume} {111}},\ \bibinfo {pages} {353} (\bibinfo {year}
  {1999})}\BibitemShut {NoStop}%
\bibitem [{\citenamefont {Gorbunova}\ and\ \citenamefont
  {Linde}(1979)}]{gorbunova1979structure}%
  \BibitemOpen
  \bibfield  {author} {\bibinfo {author} {\bibfnamefont {Y.~E.}\ \bibnamefont
  {Gorbunova}}\ and\ \bibinfo {author} {\bibfnamefont {S.}~\bibnamefont
  {Linde}},\ }\bibfield  {title} {\bibinfo {title} {Structure of crystals of
  vanadyl pyrophosphate (vo)\_2p\_2o\_7},\ }in\ \href@noop {} {\emph {\bibinfo
  {booktitle} {Doklady Akademii Nauk}}},\ Vol.\ \bibinfo {volume} {245}\
  (\bibinfo {organization} {Russian Academy of Sciences},\ \bibinfo {year}
  {1979})\ pp.\ \bibinfo {pages} {584--588}\BibitemShut {NoStop}%
\bibitem [{\citenamefont {Takahashi}\ \emph {et~al.}(2015)\citenamefont
  {Takahashi}, \citenamefont {Sugimoto}, \citenamefont {Nambu}, \citenamefont
  {Yamauchi}, \citenamefont {Hirata}, \citenamefont {Kawakami}, \citenamefont
  {Avdeev}, \citenamefont {Matsubayashi}, \citenamefont {Du}, \citenamefont
  {Kawashima} \emph {et~al.}}]{takahashi2015pressure}%
  \BibitemOpen
  \bibfield  {author} {\bibinfo {author} {\bibfnamefont {H.}~\bibnamefont
  {Takahashi}}, \bibinfo {author} {\bibfnamefont {A.}~\bibnamefont {Sugimoto}},
  \bibinfo {author} {\bibfnamefont {Y.}~\bibnamefont {Nambu}}, \bibinfo
  {author} {\bibfnamefont {T.}~\bibnamefont {Yamauchi}}, \bibinfo {author}
  {\bibfnamefont {Y.}~\bibnamefont {Hirata}}, \bibinfo {author} {\bibfnamefont
  {T.}~\bibnamefont {Kawakami}}, \bibinfo {author} {\bibfnamefont
  {M.}~\bibnamefont {Avdeev}}, \bibinfo {author} {\bibfnamefont
  {K.}~\bibnamefont {Matsubayashi}}, \bibinfo {author} {\bibfnamefont
  {F.}~\bibnamefont {Du}}, \bibinfo {author} {\bibfnamefont {C.}~\bibnamefont
  {Kawashima}}, \emph {et~al.},\ }\bibfield  {title} {\bibinfo {title}
  {Pressure-induced superconductivity in the iron-based ladder material
  bafe2s3},\ }\href@noop {} {\bibfield  {journal} {\bibinfo  {journal} {Nature
  materials}\ }\textbf {\bibinfo {volume} {14}},\ \bibinfo {pages} {1008}
  (\bibinfo {year} {2015})}\BibitemShut {NoStop}%
\bibitem [{\citenamefont {Chiari}\ \emph {et~al.}(1990)\citenamefont {Chiari},
  \citenamefont {Piovesana}, \citenamefont {Tarantelli},\ and\ \citenamefont
  {Zanazzi}}]{chiari1990exchange}%
  \BibitemOpen
  \bibfield  {author} {\bibinfo {author} {\bibfnamefont {B.}~\bibnamefont
  {Chiari}}, \bibinfo {author} {\bibfnamefont {O.}~\bibnamefont {Piovesana}},
  \bibinfo {author} {\bibfnamefont {T.}~\bibnamefont {Tarantelli}},\ and\
  \bibinfo {author} {\bibfnamefont {P.}~\bibnamefont {Zanazzi}},\ }\bibfield
  {title} {\bibinfo {title} {Exchange interaction in multinuclear transition
  metal complexes. 14. exchange interactions in a novel copper (ii)
  linear-chain compound with ladderlike structure: Cu2 (1, 4-diazacycloheptane)
  2cl4},\ }\href@noop {} {\bibfield  {journal} {\bibinfo  {journal} {Inorganic
  Chemistry}\ }\textbf {\bibinfo {volume} {29}},\ \bibinfo {pages} {1172}
  (\bibinfo {year} {1990})}\BibitemShut {NoStop}%
\bibitem [{\citenamefont {Komm}\ \emph {et~al.}(2012)\citenamefont {Komm},
  \citenamefont {Biner}, \citenamefont {Neels},\ and\ \citenamefont
  {Kr{\"a}mer}}]{komm2012c5h12n}%
  \BibitemOpen
  \bibfield  {author} {\bibinfo {author} {\bibfnamefont {T.}~\bibnamefont
  {Komm}}, \bibinfo {author} {\bibfnamefont {D.}~\bibnamefont {Biner}},
  \bibinfo {author} {\bibfnamefont {A.}~\bibnamefont {Neels}},\ and\ \bibinfo
  {author} {\bibfnamefont {K.~W.}\ \bibnamefont {Kr{\"a}mer}},\ }\bibfield
  {title} {\bibinfo {title} {(c5h12n) cu2br3: A piperidinium copper (i) bromide
  with [cu2br3]- ladders},\ }\href@noop {} {\bibfield  {journal} {\bibinfo
  {journal} {Crystals}\ }\textbf {\bibinfo {volume} {2}},\ \bibinfo {pages}
  {1434} (\bibinfo {year} {2012})}\BibitemShut {NoStop}%
\bibitem [{\citenamefont {Biswal}\ \emph {et~al.}(2023)\citenamefont {Biswal},
  \citenamefont {Guchhait}, \citenamefont {Ghosh}, \citenamefont {Sarangi},
  \citenamefont {Samal}, \citenamefont {Swain}, \citenamefont {Kumar},\ and\
  \citenamefont {Nath}}]{biswal2023crystal}%
  \BibitemOpen
  \bibfield  {author} {\bibinfo {author} {\bibfnamefont {P.}~\bibnamefont
  {Biswal}}, \bibinfo {author} {\bibfnamefont {S.}~\bibnamefont {Guchhait}},
  \bibinfo {author} {\bibfnamefont {S.}~\bibnamefont {Ghosh}}, \bibinfo
  {author} {\bibfnamefont {S.}~\bibnamefont {Sarangi}}, \bibinfo {author}
  {\bibfnamefont {D.}~\bibnamefont {Samal}}, \bibinfo {author} {\bibfnamefont
  {D.}~\bibnamefont {Swain}}, \bibinfo {author} {\bibfnamefont
  {M.}~\bibnamefont {Kumar}},\ and\ \bibinfo {author} {\bibfnamefont
  {R.}~\bibnamefont {Nath}},\ }\bibfield  {title} {\bibinfo {title} {Crystal
  structure and magnetic properties of the spin-1 2 frustrated two-leg ladder
  compounds (c 4 h 14 n 2) cu 2 x 6 (x= cl and br)},\ }\href@noop {} {\bibfield
   {journal} {\bibinfo  {journal} {Physical Review B}\ }\textbf {\bibinfo
  {volume} {108}},\ \bibinfo {pages} {134420} (\bibinfo {year}
  {2023})}\BibitemShut {NoStop}%
\bibitem [{\citenamefont {Akutagawa}\ \emph {et~al.}(2002)\citenamefont
  {Akutagawa}, \citenamefont {Hasegawa}, \citenamefont {Nakamura},\ and\
  \citenamefont {Inabe}}]{akutagawa2002supramolecular}%
  \BibitemOpen
  \bibfield  {author} {\bibinfo {author} {\bibfnamefont {T.}~\bibnamefont
  {Akutagawa}}, \bibinfo {author} {\bibfnamefont {T.}~\bibnamefont {Hasegawa}},
  \bibinfo {author} {\bibfnamefont {T.}~\bibnamefont {Nakamura}},\ and\
  \bibinfo {author} {\bibfnamefont {T.}~\bibnamefont {Inabe}},\ }\bibfield
  {title} {\bibinfo {title} {Supramolecular cation assemblies of
  hydrogen-bonded (nh4+/nh2nh3+)(crown ether) in [ni (dmit) 2]-based molecular
  conductors and magnets},\ }\href@noop {} {\bibfield  {journal} {\bibinfo
  {journal} {Journal of the American Chemical Society}\ }\textbf {\bibinfo
  {volume} {124}},\ \bibinfo {pages} {8903} (\bibinfo {year}
  {2002})}\BibitemShut {NoStop}%
\bibitem [{\citenamefont {Rovira}\ \emph {et~al.}(1997)\citenamefont {Rovira},
  \citenamefont {Veciana}, \citenamefont {Ribera}, \citenamefont {Tarr{\'e}s},
  \citenamefont {Canadell}, \citenamefont {Rousseau}, \citenamefont {Mas},
  \citenamefont {Molins}, \citenamefont {Almeida}, \citenamefont {Henriques}
  \emph {et~al.}}]{rovira1997organic}%
  \BibitemOpen
  \bibfield  {author} {\bibinfo {author} {\bibfnamefont {C.}~\bibnamefont
  {Rovira}}, \bibinfo {author} {\bibfnamefont {J.}~\bibnamefont {Veciana}},
  \bibinfo {author} {\bibfnamefont {E.}~\bibnamefont {Ribera}}, \bibinfo
  {author} {\bibfnamefont {J.}~\bibnamefont {Tarr{\'e}s}}, \bibinfo {author}
  {\bibfnamefont {E.}~\bibnamefont {Canadell}}, \bibinfo {author}
  {\bibfnamefont {R.}~\bibnamefont {Rousseau}}, \bibinfo {author}
  {\bibfnamefont {M.}~\bibnamefont {Mas}}, \bibinfo {author} {\bibfnamefont
  {E.}~\bibnamefont {Molins}}, \bibinfo {author} {\bibfnamefont
  {M.}~\bibnamefont {Almeida}}, \bibinfo {author} {\bibfnamefont {R.~T.}\
  \bibnamefont {Henriques}}, \emph {et~al.},\ }\bibfield  {title} {\bibinfo
  {title} {An organic spin-ladder molecular material},\ }\href@noop {}
  {\bibfield  {journal} {\bibinfo  {journal} {Angewandte Chemie International
  Edition in English}\ }\textbf {\bibinfo {volume} {36}},\ \bibinfo {pages}
  {2324} (\bibinfo {year} {1997})}\BibitemShut {NoStop}%
\bibitem [{\citenamefont {Nomura}\ \emph {et~al.}(2022)\citenamefont {Nomura},
  \citenamefont {Matsuda}, \citenamefont {Ikeda}, \citenamefont {Kohama},
  \citenamefont {Tsuda}, \citenamefont {Amaya}, \citenamefont {Ono},\ and\
  \citenamefont {Hosokoshi}}]{nomura2022metastable}%
  \BibitemOpen
  \bibfield  {author} {\bibinfo {author} {\bibfnamefont {K.}~\bibnamefont
  {Nomura}}, \bibinfo {author} {\bibfnamefont {Y.~H.}\ \bibnamefont {Matsuda}},
  \bibinfo {author} {\bibfnamefont {A.}~\bibnamefont {Ikeda}}, \bibinfo
  {author} {\bibfnamefont {Y.}~\bibnamefont {Kohama}}, \bibinfo {author}
  {\bibfnamefont {H.}~\bibnamefont {Tsuda}}, \bibinfo {author} {\bibfnamefont
  {N.}~\bibnamefont {Amaya}}, \bibinfo {author} {\bibfnamefont
  {T.}~\bibnamefont {Ono}},\ and\ \bibinfo {author} {\bibfnamefont
  {Y.}~\bibnamefont {Hosokoshi}},\ }\bibfield  {title} {\bibinfo {title}
  {Metastable magnetization plateaus in the s= 1 organic spin ladder bip-teno
  induced by a microsecond-pulsed megagauss field},\ }\href@noop {} {\bibfield
  {journal} {\bibinfo  {journal} {Physical Review B}\ }\textbf {\bibinfo
  {volume} {105}},\ \bibinfo {pages} {214430} (\bibinfo {year}
  {2022})}\BibitemShut {NoStop}%
\bibitem [{\citenamefont {Higashi}\ \emph {et~al.}(1997)\citenamefont
  {Higashi}, \citenamefont {Tanaka}, \citenamefont {Kobayashi}, \citenamefont
  {Ishizawa},\ and\ \citenamefont {Takami}}]{higashi1997crystal}%
  \BibitemOpen
  \bibfield  {author} {\bibinfo {author} {\bibfnamefont {I.}~\bibnamefont
  {Higashi}}, \bibinfo {author} {\bibfnamefont {T.}~\bibnamefont {Tanaka}},
  \bibinfo {author} {\bibfnamefont {K.}~\bibnamefont {Kobayashi}}, \bibinfo
  {author} {\bibfnamefont {Y.}~\bibnamefont {Ishizawa}},\ and\ \bibinfo
  {author} {\bibfnamefont {M.}~\bibnamefont {Takami}},\ }\bibfield  {title}
  {\bibinfo {title} {Crystal structure of yb41si1. 2},\ }\href@noop {}
  {\bibfield  {journal} {\bibinfo  {journal} {Journal of Solid State
  Chemistry}\ }\textbf {\bibinfo {volume} {133}},\ \bibinfo {pages} {11}
  (\bibinfo {year} {1997})}\BibitemShut {NoStop}%
\bibitem [{\citenamefont {Mori}(2006)}]{mori2006crystal}%
  \BibitemOpen
  \bibfield  {author} {\bibinfo {author} {\bibfnamefont {T.}~\bibnamefont
  {Mori}},\ }\bibfield  {title} {\bibinfo {title} {Crystal growth and magnetic
  properties of rare earth borosilicides},\ }\href@noop {} {\bibfield
  {journal} {\bibinfo  {journal} {Zeitschrift f{\"u}r
  Kristallographie-Crystalline Materials}\ }\textbf {\bibinfo {volume} {221}},\
  \bibinfo {pages} {464} (\bibinfo {year} {2006})}\BibitemShut {NoStop}%
\bibitem [{\citenamefont {Zhou}\ \emph {et~al.}(2024)\citenamefont {Zhou},
  \citenamefont {Liu}, \citenamefont {Song}, \citenamefont {Ling},
  \citenamefont {Li}, \citenamefont {Tong}, \citenamefont {Xia}, \citenamefont
  {Gong}, \citenamefont {Wang}, \citenamefont {Zhao} \emph
  {et~al.}}]{zhou2024structure}%
  \BibitemOpen
  \bibfield  {author} {\bibinfo {author} {\bibfnamefont {J.}~\bibnamefont
  {Zhou}}, \bibinfo {author} {\bibfnamefont {A.}~\bibnamefont {Liu}}, \bibinfo
  {author} {\bibfnamefont {F.}~\bibnamefont {Song}}, \bibinfo {author}
  {\bibfnamefont {L.}~\bibnamefont {Ling}}, \bibinfo {author} {\bibfnamefont
  {J.}~\bibnamefont {Li}}, \bibinfo {author} {\bibfnamefont {W.}~\bibnamefont
  {Tong}}, \bibinfo {author} {\bibfnamefont {Z.}~\bibnamefont {Xia}}, \bibinfo
  {author} {\bibfnamefont {G.}~\bibnamefont {Gong}}, \bibinfo {author}
  {\bibfnamefont {Y.}~\bibnamefont {Wang}}, \bibinfo {author} {\bibfnamefont
  {J.}~\bibnamefont {Zhao}}, \emph {et~al.},\ }\bibfield  {title} {\bibinfo
  {title} {Structure and magnetic properties of a family of two-leg spin ladder
  compounds ba2re2ge4o13 (re= pr, nd, and gd--ho)},\ }\href@noop {} {\bibfield
  {journal} {\bibinfo  {journal} {Inorganic Chemistry}\ }\textbf {\bibinfo
  {volume} {63}},\ \bibinfo {pages} {22748} (\bibinfo {year}
  {2024})}\BibitemShut {NoStop}%
\bibitem [{\citenamefont {Orlandi}\ \emph {et~al.}(2025)\citenamefont
  {Orlandi}, \citenamefont {Ciomaga~Hatnean}, \citenamefont {Mayoh},
  \citenamefont {Tidey}, \citenamefont {Riberolles}, \citenamefont
  {Balakrishnan}, \citenamefont {Manuel}, \citenamefont {Khalyavin},
  \citenamefont {Walker}, \citenamefont {Le} \emph
  {et~al.}}]{orlandi2025magnetic}%
  \BibitemOpen
  \bibfield  {author} {\bibinfo {author} {\bibfnamefont {F.}~\bibnamefont
  {Orlandi}}, \bibinfo {author} {\bibfnamefont {M.}~\bibnamefont
  {Ciomaga~Hatnean}}, \bibinfo {author} {\bibfnamefont {D.}~\bibnamefont
  {Mayoh}}, \bibinfo {author} {\bibfnamefont {J.}~\bibnamefont {Tidey}},
  \bibinfo {author} {\bibfnamefont {S.}~\bibnamefont {Riberolles}}, \bibinfo
  {author} {\bibfnamefont {G.}~\bibnamefont {Balakrishnan}}, \bibinfo {author}
  {\bibfnamefont {P.}~\bibnamefont {Manuel}}, \bibinfo {author} {\bibfnamefont
  {D.}~\bibnamefont {Khalyavin}}, \bibinfo {author} {\bibfnamefont
  {H.}~\bibnamefont {Walker}}, \bibinfo {author} {\bibfnamefont
  {M.}~\bibnamefont {Le}}, \emph {et~al.},\ }\bibfield  {title} {\bibinfo
  {title} {Magnetic properties of the zigzag ladder compound srtb 2 o 4},\
  }\href@noop {} {\bibfield  {journal} {\bibinfo  {journal} {Physical Review
  B}\ }\textbf {\bibinfo {volume} {111}},\ \bibinfo {pages} {054415} (\bibinfo
  {year} {2025})}\BibitemShut {NoStop}%
\bibitem [{\citenamefont {Canfield}\ \emph {et~al.}(2016)\citenamefont
  {Canfield}, \citenamefont {Kong}, \citenamefont {Kaluarachchi},\ and\
  \citenamefont {Jo}}]{canfield2016use}%
  \BibitemOpen
  \bibfield  {author} {\bibinfo {author} {\bibfnamefont {P.~C.}\ \bibnamefont
  {Canfield}}, \bibinfo {author} {\bibfnamefont {T.}~\bibnamefont {Kong}},
  \bibinfo {author} {\bibfnamefont {U.~S.}\ \bibnamefont {Kaluarachchi}},\ and\
  \bibinfo {author} {\bibfnamefont {N.~H.}\ \bibnamefont {Jo}},\ }\bibfield
  {title} {\bibinfo {title} {Use of frit-disc crucibles for routine and
  exploratory solution growth of single crystalline samples},\ }\href@noop {}
  {\bibfield  {journal} {\bibinfo  {journal} {Philosophical magazine}\ }\textbf
  {\bibinfo {volume} {96}},\ \bibinfo {pages} {84} (\bibinfo {year}
  {2016})}\BibitemShut {NoStop}%
\bibitem [{ESI()}]{ESI}%
  \BibitemOpen
  \href@noop {} {}\bibinfo {note} {See Supplemental Information for further
  details}\BibitemShut {NoStop}%
\bibitem [{\citenamefont {Kresse}\ and\ \citenamefont
  {Furthm{\"u}ller}(1996{\natexlab{a}})}]{kresse1996efficient}%
  \BibitemOpen
  \bibfield  {author} {\bibinfo {author} {\bibfnamefont {G.}~\bibnamefont
  {Kresse}}\ and\ \bibinfo {author} {\bibfnamefont {J.}~\bibnamefont
  {Furthm{\"u}ller}},\ }\bibfield  {title} {\bibinfo {title} {Efficient
  iterative schemes for ab initio total-energy calculations using a plane-wave
  basis set},\ }\href@noop {} {\bibfield  {journal} {\bibinfo  {journal} {Phys.
  Rev. B}\ }\textbf {\bibinfo {volume} {54}},\ \bibinfo {pages} {11169}
  (\bibinfo {year} {1996}{\natexlab{a}})}\BibitemShut {NoStop}%
\bibitem [{\citenamefont {Kresse}\ and\ \citenamefont
  {Furthm{\"u}ller}(1996{\natexlab{b}})}]{kresse1996efficiency}%
  \BibitemOpen
  \bibfield  {author} {\bibinfo {author} {\bibfnamefont {G.}~\bibnamefont
  {Kresse}}\ and\ \bibinfo {author} {\bibfnamefont {J.}~\bibnamefont
  {Furthm{\"u}ller}},\ }\bibfield  {title} {\bibinfo {title} {Efficiency of
  ab-initio total energy calculations for metals and semiconductors using a
  plane-wave basis set},\ }\href@noop {} {\bibfield  {journal} {\bibinfo
  {journal} {Comput. Mater. Sci.}\ }\textbf {\bibinfo {volume} {6}},\ \bibinfo
  {pages} {15} (\bibinfo {year} {1996}{\natexlab{b}})}\BibitemShut {NoStop}%
\bibitem [{\citenamefont {Kresse}\ and\ \citenamefont
  {Joubert}(1999)}]{kresse1999ultrasoft}%
  \BibitemOpen
  \bibfield  {author} {\bibinfo {author} {\bibfnamefont {G.}~\bibnamefont
  {Kresse}}\ and\ \bibinfo {author} {\bibfnamefont {D.}~\bibnamefont
  {Joubert}},\ }\bibfield  {title} {\bibinfo {title} {From ultrasoft
  pseudopotentials to the projector augmented-wave method},\ }\href@noop {}
  {\bibfield  {journal} {\bibinfo  {journal} {Phys. Rev. B}\ }\textbf {\bibinfo
  {volume} {59}},\ \bibinfo {pages} {1758} (\bibinfo {year}
  {1999})}\BibitemShut {NoStop}%
\bibitem [{\citenamefont {Perdew}\ \emph {et~al.}(1996)\citenamefont {Perdew},
  \citenamefont {Burke},\ and\ \citenamefont {Ernzerhof}}]{perdew1996}%
  \BibitemOpen
  \bibfield  {author} {\bibinfo {author} {\bibfnamefont {J.~P.}\ \bibnamefont
  {Perdew}}, \bibinfo {author} {\bibfnamefont {K.}~\bibnamefont {Burke}},\ and\
  \bibinfo {author} {\bibfnamefont {M.}~\bibnamefont {Ernzerhof}},\ }\bibfield
  {title} {\bibinfo {title} {Generalized gradient approximation made simple},\
  }\href@noop {} {\bibfield  {journal} {\bibinfo  {journal} {Phys. Rev. Lett.}\
  }\textbf {\bibinfo {volume} {77}},\ \bibinfo {pages} {3865} (\bibinfo {year}
  {1996})}\BibitemShut {NoStop}%
\bibitem [{\citenamefont {Hinuma}\ \emph {et~al.}(2017)\citenamefont {Hinuma},
  \citenamefont {Pizzi}, \citenamefont {Kumagai}, \citenamefont {Oba},\ and\
  \citenamefont {Tanaka}}]{hinuma2017band}%
  \BibitemOpen
  \bibfield  {author} {\bibinfo {author} {\bibfnamefont {Y.}~\bibnamefont
  {Hinuma}}, \bibinfo {author} {\bibfnamefont {G.}~\bibnamefont {Pizzi}},
  \bibinfo {author} {\bibfnamefont {Y.}~\bibnamefont {Kumagai}}, \bibinfo
  {author} {\bibfnamefont {F.}~\bibnamefont {Oba}},\ and\ \bibinfo {author}
  {\bibfnamefont {I.}~\bibnamefont {Tanaka}},\ }\bibfield  {title} {\bibinfo
  {title} {Band structure diagram paths based on crystallography},\ }\href@noop
  {} {\bibfield  {journal} {\bibinfo  {journal} {Computational Materials
  Science}\ }\textbf {\bibinfo {volume} {128}},\ \bibinfo {pages} {140}
  (\bibinfo {year} {2017})}\BibitemShut {NoStop}%
\bibitem [{\citenamefont {Togo}\ \emph {et~al.}(2024)\citenamefont {Togo},
  \citenamefont {Shinohara},\ and\ \citenamefont {Tanaka}}]{togo2024spglib}%
  \BibitemOpen
  \bibfield  {author} {\bibinfo {author} {\bibfnamefont {A.}~\bibnamefont
  {Togo}}, \bibinfo {author} {\bibfnamefont {K.}~\bibnamefont {Shinohara}},\
  and\ \bibinfo {author} {\bibfnamefont {I.}~\bibnamefont {Tanaka}},\
  }\bibfield  {title} {\bibinfo {title} {Spglib: a software library for crystal
  symmetry search},\ }\href@noop {} {\bibfield  {journal} {\bibinfo  {journal}
  {Science and Technology of Advanced Materials: Methods}\ }\textbf {\bibinfo
  {volume} {4}},\ \bibinfo {pages} {2384822} (\bibinfo {year}
  {2024})}\BibitemShut {NoStop}%
\bibitem [{\citenamefont {Ortiz}\ \emph
  {et~al.}(2023{\natexlab{a}})\citenamefont {Ortiz}, \citenamefont {Pokharel},
  \citenamefont {Gundayao}, \citenamefont {Li}, \citenamefont {Kaboudvand},
  \citenamefont {Kautzsch}, \citenamefont {Sarker}, \citenamefont {Ruff},
  \citenamefont {Hogan}, \citenamefont {Alvarado} \emph
  {et~al.}}]{ortiz2023ybv}%
  \BibitemOpen
  \bibfield  {author} {\bibinfo {author} {\bibfnamefont {B.~R.}\ \bibnamefont
  {Ortiz}}, \bibinfo {author} {\bibfnamefont {G.}~\bibnamefont {Pokharel}},
  \bibinfo {author} {\bibfnamefont {M.}~\bibnamefont {Gundayao}}, \bibinfo
  {author} {\bibfnamefont {H.}~\bibnamefont {Li}}, \bibinfo {author}
  {\bibfnamefont {F.}~\bibnamefont {Kaboudvand}}, \bibinfo {author}
  {\bibfnamefont {L.}~\bibnamefont {Kautzsch}}, \bibinfo {author}
  {\bibfnamefont {S.}~\bibnamefont {Sarker}}, \bibinfo {author} {\bibfnamefont
  {J.~P.}\ \bibnamefont {Ruff}}, \bibinfo {author} {\bibfnamefont
  {T.}~\bibnamefont {Hogan}}, \bibinfo {author} {\bibfnamefont {S.~J.~G.}\
  \bibnamefont {Alvarado}}, \emph {et~al.},\ }\bibfield  {title} {\bibinfo
  {title} {{YbV$_3$Sb$_4$ and EuV$_3$Sb$_4$ vanadium-based kagome metals with
  Yb$^{2+}$ and Eu$^{2+}$ zigzag chains}},\ }\href@noop {} {\bibfield
  {journal} {\bibinfo  {journal} {Phys. Rev. Mater.}\ }\textbf {\bibinfo
  {volume} {7}},\ \bibinfo {pages} {064201} (\bibinfo {year}
  {2023}{\natexlab{a}})}\BibitemShut {NoStop}%
\bibitem [{\citenamefont {Ortiz}\ \emph {et~al.}(2019)\citenamefont {Ortiz},
  \citenamefont {Gomes}, \citenamefont {Morey}, \citenamefont {Winiarski},
  \citenamefont {Bordelon}, \citenamefont {Mangum}, \citenamefont {Oswald},
  \citenamefont {Rodriguez-Rivera}, \citenamefont {Neilson}, \citenamefont
  {Wilson} \emph {et~al.}}]{ortiz2019new}%
  \BibitemOpen
  \bibfield  {author} {\bibinfo {author} {\bibfnamefont {B.~R.}\ \bibnamefont
  {Ortiz}}, \bibinfo {author} {\bibfnamefont {L.~C.}\ \bibnamefont {Gomes}},
  \bibinfo {author} {\bibfnamefont {J.~R.}\ \bibnamefont {Morey}}, \bibinfo
  {author} {\bibfnamefont {M.}~\bibnamefont {Winiarski}}, \bibinfo {author}
  {\bibfnamefont {M.}~\bibnamefont {Bordelon}}, \bibinfo {author}
  {\bibfnamefont {J.~S.}\ \bibnamefont {Mangum}}, \bibinfo {author}
  {\bibfnamefont {I.~W.}\ \bibnamefont {Oswald}}, \bibinfo {author}
  {\bibfnamefont {J.~A.}\ \bibnamefont {Rodriguez-Rivera}}, \bibinfo {author}
  {\bibfnamefont {J.~R.}\ \bibnamefont {Neilson}}, \bibinfo {author}
  {\bibfnamefont {S.~D.}\ \bibnamefont {Wilson}}, \emph {et~al.},\ }\bibfield
  {title} {\bibinfo {title} {{New kagome prototype materials: discovery of
  KV$_3$Sb$_5$, RbV$_3$Sb$_5$, and CsV$_3$Sb$_5$}},\ }\href@noop {} {\bibfield
  {journal} {\bibinfo  {journal} {Phys. Rev. Materials}\ }\textbf {\bibinfo
  {volume} {3}},\ \bibinfo {pages} {094407} (\bibinfo {year}
  {2019})}\BibitemShut {NoStop}%
\bibitem [{\citenamefont {Ortiz}\ \emph
  {et~al.}(2023{\natexlab{b}})\citenamefont {Ortiz}, \citenamefont {Miao},
  \citenamefont {Parker}, \citenamefont {Yang}, \citenamefont {Samolyuk},
  \citenamefont {Clements}, \citenamefont {Rajapitamahuni}, \citenamefont
  {Yilmaz}, \citenamefont {Vescovo}, \citenamefont {Yan} \emph
  {et~al.}}]{ortiz2023evolution}%
  \BibitemOpen
  \bibfield  {author} {\bibinfo {author} {\bibfnamefont {B.~R.}\ \bibnamefont
  {Ortiz}}, \bibinfo {author} {\bibfnamefont {H.}~\bibnamefont {Miao}},
  \bibinfo {author} {\bibfnamefont {D.~S.}\ \bibnamefont {Parker}}, \bibinfo
  {author} {\bibfnamefont {F.}~\bibnamefont {Yang}}, \bibinfo {author}
  {\bibfnamefont {G.~D.}\ \bibnamefont {Samolyuk}}, \bibinfo {author}
  {\bibfnamefont {E.~M.}\ \bibnamefont {Clements}}, \bibinfo {author}
  {\bibfnamefont {A.}~\bibnamefont {Rajapitamahuni}}, \bibinfo {author}
  {\bibfnamefont {T.}~\bibnamefont {Yilmaz}}, \bibinfo {author} {\bibfnamefont
  {E.}~\bibnamefont {Vescovo}}, \bibinfo {author} {\bibfnamefont
  {J.}~\bibnamefont {Yan}}, \emph {et~al.},\ }\bibfield  {title} {\bibinfo
  {title} {{Evolution of Highly Anisotropic Magnetism in the Titanium-Based
  Kagome Metals LnTi$_3$Bi$_4$ (Ln:
  La{\textperiodcentered}{\textperiodcentered}{\textperiodcentered} Gd$^{3+}$,
  Eu$^{2+}$, Yb$^{2+}$)}},\ }\href@noop {} {\bibfield  {journal} {\bibinfo
  {journal} {Chemistry of Materials}\ }\textbf {\bibinfo {volume} {35}},\
  \bibinfo {pages} {9756} (\bibinfo {year} {2023}{\natexlab{b}})}\BibitemShut
  {NoStop}%
\bibitem [{\citenamefont {Ovchinnikov}\ and\ \citenamefont
  {Bobev}(2018)}]{ovchinnikov2018synthesis}%
  \BibitemOpen
  \bibfield  {author} {\bibinfo {author} {\bibfnamefont {A.}~\bibnamefont
  {Ovchinnikov}}\ and\ \bibinfo {author} {\bibfnamefont {S.}~\bibnamefont
  {Bobev}},\ }\bibfield  {title} {\bibinfo {title} {{Synthesis, Crystal and
  Electronic Structure of the Titanium Bismuthides Sr$_5$Ti$_{12}$Bi$_{19+x}$,
  Ba$_5$Ti$_{12}$Bi$_{19+x}$, and
  Sr$_{5-\delta}$Eu$_\delta$Ti$_{12}$Bi$_{19+x}$ (x=0.5--1.0; $\delta$=2.4,
  4.0)}},\ }\href@noop {} {\bibfield  {journal} {\bibinfo  {journal} {Eur. J.
  Inorg. Chem.}\ }\textbf {\bibinfo {volume} {2018}},\ \bibinfo {pages} {1266}
  (\bibinfo {year} {2018})}\BibitemShut {NoStop}%
\bibitem [{\citenamefont {Ovchinnikov}\ and\ \citenamefont
  {Bobev}(2019)}]{ovchinnikov2019bismuth}%
  \BibitemOpen
  \bibfield  {author} {\bibinfo {author} {\bibfnamefont {A.}~\bibnamefont
  {Ovchinnikov}}\ and\ \bibinfo {author} {\bibfnamefont {S.}~\bibnamefont
  {Bobev}},\ }\bibfield  {title} {\bibinfo {title} {Bismuth as a reactive
  solvent in the synthesis of multicomponent transition-metal-bearing
  bismuthides},\ }\href@noop {} {\bibfield  {journal} {\bibinfo  {journal}
  {Inorg. Chem.}\ }\textbf {\bibinfo {volume} {59}},\ \bibinfo {pages} {3459}
  (\bibinfo {year} {2019})}\BibitemShut {NoStop}%
\bibitem [{\citenamefont {Motoyama}\ \emph {et~al.}(2018)\citenamefont
  {Motoyama}, \citenamefont {Sezaki}, \citenamefont {Gouchi}, \citenamefont
  {Miyoshi}, \citenamefont {Nishigori}, \citenamefont {Mutou}, \citenamefont
  {Fujiwara},\ and\ \citenamefont {Uwatoko}}]{motoyama2018magnetic}%
  \BibitemOpen
  \bibfield  {author} {\bibinfo {author} {\bibfnamefont {G.}~\bibnamefont
  {Motoyama}}, \bibinfo {author} {\bibfnamefont {M.}~\bibnamefont {Sezaki}},
  \bibinfo {author} {\bibfnamefont {J.}~\bibnamefont {Gouchi}}, \bibinfo
  {author} {\bibfnamefont {K.}~\bibnamefont {Miyoshi}}, \bibinfo {author}
  {\bibfnamefont {S.}~\bibnamefont {Nishigori}}, \bibinfo {author}
  {\bibfnamefont {T.}~\bibnamefont {Mutou}}, \bibinfo {author} {\bibfnamefont
  {K.}~\bibnamefont {Fujiwara}},\ and\ \bibinfo {author} {\bibfnamefont
  {Y.}~\bibnamefont {Uwatoko}},\ }\bibfield  {title} {\bibinfo {title}
  {{Magnetic properties of new antiferromagnetic heavy-fermion compounds,
  Ce$_3$TiBi$_5$ and CeTi$_3$Bi$_4$}},\ }\href@noop {} {\bibfield  {journal}
  {\bibinfo  {journal} {Physica B Condens.}\ }\textbf {\bibinfo {volume}
  {536}},\ \bibinfo {pages} {142} (\bibinfo {year} {2018})}\BibitemShut
  {NoStop}%
\bibitem [{\citenamefont {Chen}\ \emph {et~al.}(2024)\citenamefont {Chen},
  \citenamefont {Zhou}, \citenamefont {Zhang}, \citenamefont {Ji},
  \citenamefont {Liao}, \citenamefont {Ji}, \citenamefont {Li}, \citenamefont
  {Guo}, \citenamefont {Shen}, \citenamefont {Yu} \emph
  {et~al.}}]{chen2023134}%
  \BibitemOpen
  \bibfield  {author} {\bibinfo {author} {\bibfnamefont {L.}~\bibnamefont
  {Chen}}, \bibinfo {author} {\bibfnamefont {Y.}~\bibnamefont {Zhou}}, \bibinfo
  {author} {\bibfnamefont {H.}~\bibnamefont {Zhang}}, \bibinfo {author}
  {\bibfnamefont {X.}~\bibnamefont {Ji}}, \bibinfo {author} {\bibfnamefont
  {K.}~\bibnamefont {Liao}}, \bibinfo {author} {\bibfnamefont {Y.}~\bibnamefont
  {Ji}}, \bibinfo {author} {\bibfnamefont {Y.}~\bibnamefont {Li}}, \bibinfo
  {author} {\bibfnamefont {Z.}~\bibnamefont {Guo}}, \bibinfo {author}
  {\bibfnamefont {X.}~\bibnamefont {Shen}}, \bibinfo {author} {\bibfnamefont
  {R.}~\bibnamefont {Yu}}, \emph {et~al.},\ }\bibfield  {title} {\bibinfo
  {title} {Tunable magnetism in titanium-based kagome metals by rare-earth
  engineering and high pressure},\ }\href@noop {} {\bibfield  {journal}
  {\bibinfo  {journal} {Communications Materials}\ }\textbf {\bibinfo {volume}
  {5}},\ \bibinfo {pages} {73} (\bibinfo {year} {2024})}\BibitemShut {NoStop}%
\bibitem [{\citenamefont {Guo}\ \emph {et~al.}({\natexlab{a}})\citenamefont
  {Guo}, \citenamefont {Zhou}, \citenamefont {Ding}, \citenamefont {Qu},
  \citenamefont {Liu}, \citenamefont {Du}, \citenamefont {Zhang}, \citenamefont
  {Li}, \citenamefont {Zhang}, \citenamefont {Zhou}, \citenamefont {Qi},
  \citenamefont {Guo}, \citenamefont {Wang}, \citenamefont {Fei}, \citenamefont
  {Huang}, \citenamefont {Qian}, \citenamefont {Shen}, \citenamefont {Weng},\
  and\ \citenamefont {Song}}]{guo2023134}%
  \BibitemOpen
  \bibfield  {author} {\bibinfo {author} {\bibfnamefont {J.}~\bibnamefont
  {Guo}}, \bibinfo {author} {\bibfnamefont {L.}~\bibnamefont {Zhou}}, \bibinfo
  {author} {\bibfnamefont {J.}~\bibnamefont {Ding}}, \bibinfo {author}
  {\bibfnamefont {G.}~\bibnamefont {Qu}}, \bibinfo {author} {\bibfnamefont
  {Z.}~\bibnamefont {Liu}}, \bibinfo {author} {\bibfnamefont {Y.}~\bibnamefont
  {Du}}, \bibinfo {author} {\bibfnamefont {H.}~\bibnamefont {Zhang}}, \bibinfo
  {author} {\bibfnamefont {J.}~\bibnamefont {Li}}, \bibinfo {author}
  {\bibfnamefont {Y.}~\bibnamefont {Zhang}}, \bibinfo {author} {\bibfnamefont
  {F.}~\bibnamefont {Zhou}}, \bibinfo {author} {\bibfnamefont {W.}~\bibnamefont
  {Qi}}, \bibinfo {author} {\bibfnamefont {F.}~\bibnamefont {Guo}}, \bibinfo
  {author} {\bibfnamefont {T.}~\bibnamefont {Wang}}, \bibinfo {author}
  {\bibfnamefont {F.}~\bibnamefont {Fei}}, \bibinfo {author} {\bibfnamefont
  {Y.}~\bibnamefont {Huang}}, \bibinfo {author} {\bibfnamefont
  {T.}~\bibnamefont {Qian}}, \bibinfo {author} {\bibfnamefont {D.}~\bibnamefont
  {Shen}}, \bibinfo {author} {\bibfnamefont {H.}~\bibnamefont {Weng}},\ and\
  \bibinfo {author} {\bibfnamefont {F.}~\bibnamefont {Song}},\ }\bibfield
  {title} {\bibinfo {title} {{Magnetic kagome materials RETi$_3$Bi$_4$ family
  with weak interlayer interactions}},\ }\href@noop {} {\bibfield  {journal}
  {\bibinfo  {journal} {\textbf{2023}, arXiv:2308.14509v1 [cond-mat.mtrl-sci].
  arXiv.org e-Print archive. https://arxiv.org/pdf/2308.14509.pdf (Accessed
  10-24-2023)}\ } ({\natexlab{a}})}\BibitemShut {NoStop}%
\bibitem [{\citenamefont {Ortiz}\ \emph {et~al.}(2024)\citenamefont {Ortiz},
  \citenamefont {Zhang}, \citenamefont {G{\'o}rnicka}, \citenamefont {Parker},
  \citenamefont {Samolyuk}, \citenamefont {Yang}, \citenamefont {Miao},
  \citenamefont {Lu}, \citenamefont {Moore}, \citenamefont {May} \emph
  {et~al.}}]{ortiz2024intricate}%
  \BibitemOpen
  \bibfield  {author} {\bibinfo {author} {\bibfnamefont {B.~R.}\ \bibnamefont
  {Ortiz}}, \bibinfo {author} {\bibfnamefont {H.}~\bibnamefont {Zhang}},
  \bibinfo {author} {\bibfnamefont {K.}~\bibnamefont {G{\'o}rnicka}}, \bibinfo
  {author} {\bibfnamefont {D.~S.}\ \bibnamefont {Parker}}, \bibinfo {author}
  {\bibfnamefont {G.~D.}\ \bibnamefont {Samolyuk}}, \bibinfo {author}
  {\bibfnamefont {F.}~\bibnamefont {Yang}}, \bibinfo {author} {\bibfnamefont
  {H.}~\bibnamefont {Miao}}, \bibinfo {author} {\bibfnamefont {Q.}~\bibnamefont
  {Lu}}, \bibinfo {author} {\bibfnamefont {R.~G.}\ \bibnamefont {Moore}},
  \bibinfo {author} {\bibfnamefont {A.~F.}\ \bibnamefont {May}}, \emph
  {et~al.},\ }\bibfield  {title} {\bibinfo {title} {Intricate magnetic
  landscape in antiferromagnetic kagome metal tbti3bi4 and interplay with
  ln2--x ti6+ x bi9 (ln:
  Tb{\textperiodcentered}{\textperiodcentered}{\textperiodcentered} lu)
  shurikagome metals},\ }\href@noop {} {\bibfield  {journal} {\bibinfo
  {journal} {Chemistry of Materials}\ }\textbf {\bibinfo {volume} {36}},\
  \bibinfo {pages} {8002} (\bibinfo {year} {2024})}\BibitemShut {NoStop}%
\bibitem [{\citenamefont {Guo}\ \emph {et~al.}({\natexlab{b}})\citenamefont
  {Guo}, \citenamefont {Ma}, \citenamefont {Liu}, \citenamefont {Wu},
  \citenamefont {Wang}, \citenamefont {Shi}, \citenamefont {Li},\ and\
  \citenamefont {Jia}}]{guo20241}%
  \BibitemOpen
  \bibfield  {author} {\bibinfo {author} {\bibfnamefont {K.}~\bibnamefont
  {Guo}}, \bibinfo {author} {\bibfnamefont {Z.}~\bibnamefont {Ma}}, \bibinfo
  {author} {\bibfnamefont {H.}~\bibnamefont {Liu}}, \bibinfo {author}
  {\bibfnamefont {Z.}~\bibnamefont {Wu}}, \bibinfo {author} {\bibfnamefont
  {J.}~\bibnamefont {Wang}}, \bibinfo {author} {\bibfnamefont {Y.}~\bibnamefont
  {Shi}}, \bibinfo {author} {\bibfnamefont {Y.}~\bibnamefont {Li}},\ and\
  \bibinfo {author} {\bibfnamefont {S.}~\bibnamefont {Jia}},\ }\bibfield
  {title} {\bibinfo {title} {{1/3 and other magnetization plateaus in a
  quasi-one-dimensional Ising magnet TbTi$_3$Bi$_4$ with zigzag spin chain}},\
  }\href@noop {} {\bibfield  {journal} {\bibinfo  {journal} {\textbf{2024},
  arXiv:2405.09280v1 [cond-mat.str-el]. arXiv.org e-Print archive.
  https://arxiv.org/pdf/2405.09280v1 (Accessed 6-25-2024)}\ }
  ({\natexlab{b}})}\BibitemShut {NoStop}%
\bibitem [{\citenamefont {Cheng}\ \emph {et~al.}()\citenamefont {Cheng},
  \citenamefont {Wang}, \citenamefont {Nie}, \citenamefont {Ying},
  \citenamefont {Li}, \citenamefont {Li}, \citenamefont {Xu}, \citenamefont
  {Chen}, \citenamefont {Koban}, \citenamefont {Borrmann} \emph
  {et~al.}}]{cheng2024giant}%
  \BibitemOpen
  \bibfield  {author} {\bibinfo {author} {\bibfnamefont {E.}~\bibnamefont
  {Cheng}}, \bibinfo {author} {\bibfnamefont {K.}~\bibnamefont {Wang}},
  \bibinfo {author} {\bibfnamefont {S.}~\bibnamefont {Nie}}, \bibinfo {author}
  {\bibfnamefont {T.}~\bibnamefont {Ying}}, \bibinfo {author} {\bibfnamefont
  {Z.}~\bibnamefont {Li}}, \bibinfo {author} {\bibfnamefont {Y.}~\bibnamefont
  {Li}}, \bibinfo {author} {\bibfnamefont {Y.}~\bibnamefont {Xu}}, \bibinfo
  {author} {\bibfnamefont {H.}~\bibnamefont {Chen}}, \bibinfo {author}
  {\bibfnamefont {R.}~\bibnamefont {Koban}}, \bibinfo {author} {\bibfnamefont
  {H.}~\bibnamefont {Borrmann}}, \emph {et~al.},\ }\bibfield  {title} {\bibinfo
  {title} {Giant anomalous hall effect and band folding in a kagome metal with
  mixed dimensionality},\ }\href@noop {} {\bibinfo  {journal} {\textbf{2024},
  arXiv:2405.16831v1 [cond-mat.str-el]. arXiv.org e-Print archive.
  https://arxiv.org/pdf/2405.16831v1 (Accessed 6-25-2024)}\ }\BibitemShut
  {NoStop}%
\bibitem [{\citenamefont {Bie}\ \emph {et~al.}(2007)\citenamefont {Bie},
  \citenamefont {Moore}, \citenamefont {Piercey}, \citenamefont {Tkachuk},
  \citenamefont {Zelinska},\ and\ \citenamefont {Mar}}]{bie2007ternary}%
  \BibitemOpen
\bibfield  {journal} {  }\bibfield  {author} {\bibinfo {author} {\bibfnamefont
  {H.}~\bibnamefont {Bie}}, \bibinfo {author} {\bibfnamefont {S.~D.}\
  \bibnamefont {Moore}}, \bibinfo {author} {\bibfnamefont {D.~G.}\ \bibnamefont
  {Piercey}}, \bibinfo {author} {\bibfnamefont {A.~V.}\ \bibnamefont
  {Tkachuk}}, \bibinfo {author} {\bibfnamefont {O.~Y.}\ \bibnamefont
  {Zelinska}},\ and\ \bibinfo {author} {\bibfnamefont {A.}~\bibnamefont
  {Mar}},\ }\bibfield  {title} {\bibinfo {title} {{Ternary rare-earth titanium
  antimonides: phase equilibria in the RE--Ti--Sb (RE= La, Er) systems and
  crystal structures of RE$_2$Ti$_7$Sb$_{12}$ (RE= La, Ce, Pr, Nd) and
  RETi$_3$(Sn$_x$Sb$_{1-x}$)$_4$ (RE= Nd, Sm)}},\ }\href@noop {} {\bibfield
  {journal} {\bibinfo  {journal} {J. Solid State Chem."}\ }\textbf {\bibinfo
  {volume} {180}},\ \bibinfo {pages} {2216} (\bibinfo {year}
  {2007})}\BibitemShut {NoStop}%
\bibitem [{\citenamefont {Zeng}\ \emph {et~al.}(2024)\citenamefont {Zeng},
  \citenamefont {Ge}, \citenamefont {Wu}, \citenamefont {Ruan}, \citenamefont
  {Li}, \citenamefont {Wang}, \citenamefont {Li}, \citenamefont {Ling},
  \citenamefont {Tong}, \citenamefont {Huang} \emph
  {et~al.}}]{zeng2024k2renb5o15}%
  \BibitemOpen
  \bibfield  {author} {\bibinfo {author} {\bibfnamefont {Q.}~\bibnamefont
  {Zeng}}, \bibinfo {author} {\bibfnamefont {H.}~\bibnamefont {Ge}}, \bibinfo
  {author} {\bibfnamefont {M.}~\bibnamefont {Wu}}, \bibinfo {author}
  {\bibfnamefont {S.}~\bibnamefont {Ruan}}, \bibinfo {author} {\bibfnamefont
  {T.}~\bibnamefont {Li}}, \bibinfo {author} {\bibfnamefont {Z.}~\bibnamefont
  {Wang}}, \bibinfo {author} {\bibfnamefont {J.}~\bibnamefont {Li}}, \bibinfo
  {author} {\bibfnamefont {L.}~\bibnamefont {Ling}}, \bibinfo {author}
  {\bibfnamefont {W.}~\bibnamefont {Tong}}, \bibinfo {author} {\bibfnamefont
  {S.}~\bibnamefont {Huang}}, \emph {et~al.},\ }\bibfield  {title} {\bibinfo
  {title} {K2renb5o15 (re= ce, pr, nd, sm, gd--ho): A family of
  quasi-one-dimensional spin-chain compounds with large interchain distance},\
  }\href@noop {} {\bibfield  {journal} {\bibinfo  {journal} {Chemistry of
  Materials}\ }\textbf {\bibinfo {volume} {36}},\ \bibinfo {pages} {2867}
  (\bibinfo {year} {2024})}\BibitemShut {NoStop}%
\bibitem [{\citenamefont {Algra}\ \emph {et~al.}(1978)\citenamefont {Algra},
  \citenamefont {De~Jongh},\ and\ \citenamefont {Carlin}}]{algra1978one}%
  \BibitemOpen
  \bibfield  {author} {\bibinfo {author} {\bibfnamefont {H.}~\bibnamefont
  {Algra}}, \bibinfo {author} {\bibfnamefont {L.}~\bibnamefont {De~Jongh}},\
  and\ \bibinfo {author} {\bibfnamefont {R.}~\bibnamefont {Carlin}},\
  }\bibfield  {title} {\bibinfo {title} {One-and two-dimensional s= 12
  heisenberg antiferromagnetism in cu (c5h5no) 6 (clo4) 2 and cu (c5h5no) 6
  (bf4) 2, respectively},\ }\href@noop {} {\bibfield  {journal} {\bibinfo
  {journal} {Physica B+ C}\ }\textbf {\bibinfo {volume} {93}},\ \bibinfo
  {pages} {24} (\bibinfo {year} {1978})}\BibitemShut {NoStop}%
\bibitem [{\citenamefont {Sologubenko}\ \emph {et~al.}(2007)\citenamefont
  {Sologubenko}, \citenamefont {Lorenz}, \citenamefont {Ott},\ and\
  \citenamefont {Freimuth}}]{sologubenko2007thermal}%
  \BibitemOpen
  \bibfield  {author} {\bibinfo {author} {\bibfnamefont {A.}~\bibnamefont
  {Sologubenko}}, \bibinfo {author} {\bibfnamefont {T.}~\bibnamefont {Lorenz}},
  \bibinfo {author} {\bibfnamefont {H.~R.}\ \bibnamefont {Ott}},\ and\ \bibinfo
  {author} {\bibfnamefont {A.}~\bibnamefont {Freimuth}},\ }\bibfield  {title}
  {\bibinfo {title} {Thermal conductivity via magnetic excitations in
  spin-chain materials},\ }\href@noop {} {\bibfield  {journal} {\bibinfo
  {journal} {Journal of Low Temperature Physics}\ }\textbf {\bibinfo {volume}
  {147}},\ \bibinfo {pages} {387} (\bibinfo {year} {2007})}\BibitemShut
  {NoStop}%
\bibitem [{\citenamefont {Sologubenko}\ \emph {et~al.}(2001)\citenamefont
  {Sologubenko}, \citenamefont {Gianno}, \citenamefont {Ott}, \citenamefont
  {Vietkine},\ and\ \citenamefont {Revcolevschi}}]{sologubenko2001heat}%
  \BibitemOpen
  \bibfield  {author} {\bibinfo {author} {\bibfnamefont {A.}~\bibnamefont
  {Sologubenko}}, \bibinfo {author} {\bibfnamefont {K.}~\bibnamefont {Gianno}},
  \bibinfo {author} {\bibfnamefont {H.}~\bibnamefont {Ott}}, \bibinfo {author}
  {\bibfnamefont {A.}~\bibnamefont {Vietkine}},\ and\ \bibinfo {author}
  {\bibfnamefont {A.}~\bibnamefont {Revcolevschi}},\ }\bibfield  {title}
  {\bibinfo {title} {Heat transport by lattice and spin excitations in the
  spin-chain compounds srcuo 2 and sr 2 cuo 3},\ }\href@noop {} {\bibfield
  {journal} {\bibinfo  {journal} {Physical Review B}\ }\textbf {\bibinfo
  {volume} {64}},\ \bibinfo {pages} {054412} (\bibinfo {year}
  {2001})}\BibitemShut {NoStop}%
\bibitem [{\citenamefont {Chernyshev}\ and\ \citenamefont
  {Rozhkov}(2005)}]{chernyshev2005thermal}%
  \BibitemOpen
  \bibfield  {author} {\bibinfo {author} {\bibfnamefont {A.}~\bibnamefont
  {Chernyshev}}\ and\ \bibinfo {author} {\bibfnamefont {A.}~\bibnamefont
  {Rozhkov}},\ }\bibfield  {title} {\bibinfo {title} {Thermal transport in
  antiferromagnetic spin-chain materials},\ }\href@noop {} {\bibfield
  {journal} {\bibinfo  {journal} {Physical Review B—Condensed Matter and
  Materials Physics}\ }\textbf {\bibinfo {volume} {72}},\ \bibinfo {pages}
  {104423} (\bibinfo {year} {2005})}\BibitemShut {NoStop}%
\bibitem [{\citenamefont {Kl{\"u}mper}\ and\ \citenamefont
  {Sakai}(2002)}]{klumper2002thermal}%
  \BibitemOpen
  \bibfield  {author} {\bibinfo {author} {\bibfnamefont {A.}~\bibnamefont
  {Kl{\"u}mper}}\ and\ \bibinfo {author} {\bibfnamefont {K.}~\bibnamefont
  {Sakai}},\ }\bibfield  {title} {\bibinfo {title} {The thermal conductivity of
  the spin-$1/2$ xxz chain at arbitrary temperature},\ }\href@noop {}
  {\bibfield  {journal} {\bibinfo  {journal} {Journal of Physics A:
  Mathematical and General}\ }\textbf {\bibinfo {volume} {35}},\ \bibinfo
  {pages} {2173} (\bibinfo {year} {2002})}\BibitemShut {NoStop}%
\bibitem [{\citenamefont {Sologubenko}\ \emph {et~al.}(2000)\citenamefont
  {Sologubenko}, \citenamefont {Felder}, \citenamefont {Gianno}, \citenamefont
  {Ott}, \citenamefont {Vietkine},\ and\ \citenamefont
  {Revcolevschi}}]{sologubenko2000thermal}%
  \BibitemOpen
  \bibfield  {author} {\bibinfo {author} {\bibfnamefont {A.}~\bibnamefont
  {Sologubenko}}, \bibinfo {author} {\bibfnamefont {E.}~\bibnamefont {Felder}},
  \bibinfo {author} {\bibfnamefont {K.}~\bibnamefont {Gianno}}, \bibinfo
  {author} {\bibfnamefont {H.}~\bibnamefont {Ott}}, \bibinfo {author}
  {\bibfnamefont {A.}~\bibnamefont {Vietkine}},\ and\ \bibinfo {author}
  {\bibfnamefont {A.}~\bibnamefont {Revcolevschi}},\ }\bibfield  {title}
  {\bibinfo {title} {Thermal conductivity and specific heat of the linear chain
  cuprate sr 2 cuo 3: Evidence for thermal transport via spinons},\ }\href@noop
  {} {\bibfield  {journal} {\bibinfo  {journal} {Physical Review B}\ }\textbf
  {\bibinfo {volume} {62}},\ \bibinfo {pages} {R6108} (\bibinfo {year}
  {2000})}\BibitemShut {NoStop}%
\end{thebibliography}%

\end{document}